\documentclass[english,showpacs,preprint,superscriptaddress]{revtex4-1}
\usepackage{lmodern}
\usepackage{lmodern}
\usepackage[T1]{fontenc}
\usepackage{babel}
\usepackage{mathtools}
\usepackage{amsmath}
\usepackage{amssymb}
\usepackage[unicode=true,
 bookmarks=false,
 breaklinks=false,pdfborder={0 0 1},backref=false,colorlinks=false]
 {hyperref}

\makeatletter

\usepackage{amsthm}\usepackage{latexsym}\usepackage{bm}\usepackage{amsfonts}

\usepackage{babel}

\usepackage{babel}

\usepackage{babel}

\makeatother

\begin{document}
\title{General field theory and weak Euler-Lagrange equation for classical
particle-field systems in plasma physics}
\author{Peifeng Fan}
\affiliation{Institute of Plasma Physics, Chinese Academy of Sciences, Hefei, Anhui
230031, China }
\affiliation{Science Island Branch of Graduate School, University of Science and
Technology of China, Hefei, Anhui 230031, China}
\author{Hong Qin }
\email{corresponding author: hongqin@princeton.edu}

\affiliation{Princeton Plasma Physics Laboratory, Princeton University, Princeton,
NJ 08543, USA}
\affiliation{School of Physical Sciences, University of Science and Technology
of China, Hefei, Anhui 230026, China}
\author{Jianyuan Xiao}
\affiliation{School of Physical Sciences, University of Science and Technology
of China, Hefei, Anhui 230026, China}
\author{Nong Xiang}
\affiliation{Institute of Plasma Physics, Chinese Academy of Sciences, Hefei, Anhui
230031, China }
\begin{abstract}
A general field theory for classical particle-field systems is developed.
Compared with the standard classical field theory, the distinguish
feature of a classical particle-field system is that the particles
and fields reside on different manifolds. The fields are defined on
the 4D space-time, whereas each particle's trajectory is defined on
the 1D time-axis. As a consequence, the standard Noether's procedure
for deriving local conservation laws in space-time from symmetries
is not applicable without modification. To overcome this difficulty,
a weak Euler-Lagrange equation for particles is developed on the 4D
space-time, which plays a pivotal role in establishing the connections
between symmetries and local conservation laws in space-time. Especially,
the non-vanishing Euler derivative in the weak Euler-Lagrangian equation
generates a new current in the conservation laws. Several examples
from plasma physics are studied as special cases of the general field
theory. In particular, the relations between the rotational symmetry
and angular momentum conservation for the Klimontovich-Poisson system
and the Klimontovich-Darwin system are established. 
\end{abstract}
\pacs{52.35.Hr, 52.35.-g, 52.35.We, 42.50.Tx, 52.50.Sw}
\maketitle

\section{Introduction}

It has been widely accepted as a fundamental principle of physics
that conservation laws of particle systems or field systems can be
derived from the symmetries that the systems admit. This is the well-known
Noether's theorem \citep{Noether1918}.

Classical particle-field systems, where many particles evolve under
self-generated interacting fields, are often encountered in plasma
physics \citep{Low1958,Newcomb1962,Ichimaru1973,Littlejohn1983,Ye1992,Sugama2000,Qin2007,Scott2010,Sugama2013,Sugama2018},
astrophysics \citep{Uchida1997a,Uchida1997b,Andreasson2011,Lehner2012,Contopoulos2013,Brennan2014,Dvornikov2018},
and accelerator physics \citep{Davison2001,Qin2010}. For classical
systems with particles and self-generated interacting fields, the
connections between conservation laws and symmetries have been established
only recently \citep{Landau1975,Qin2014,Fan2018}. It was pointed
out \citep{Qin2014,Fan2018} that the standard Euler-Lagrange (EL)
equation for particles are not applicable in Noether's procedure,
because the dynamics of particles and fields are defined on manifolds
with different dimensions. Instead, a weak EL equation for particles
should be used to establish the link between the conservation laws
and symmetries.

The systems discussed in \citep{Qin2014,Fan2018} are some special
particle-field systems such as the Klimontovich-Poisson (KP) system,
the Klimontovich-Darwin (KD) system and the Klimontovich-Maxwell (KM)
system. And only a special symmetry, i.e., the space-time translation
symmetry, is considered. In this study, we extend the theory to general
symmetries in general particle-field systems. The generalized theory
can be also viewed as a generalized version of Noether's theorem for
systems with classical particles and fields residing on different
manifolds.

As special cases and applications of the general theory, we study
the time translation symmetry of the KP system and the rotational
symmetry for the KP and KD systems. The energy conservation law of
the KP system, as a result of the time translation symmetry, agrees
with the result of Ref.$\thinspace$\citep{Qin2014}. The relations
between the rotational symmetry and angular momentum conservation
for the KP and KD systems are established. In this case, the rotation
of the vector potential for the KD system needs to be included as
a part of the symmetry that the system admits. Without the rotation
of the vector potential, the rotation of the position alone does not
preserve the Lagrangian. Of course, the rotation of the vector potential
is the representation of the rotational symmetry in the fiber of the
vector bundle at each space-time location.

It is also worth noting that the theory we develop here is not limited
to study the KP, KD and some special systems, it can be used for general
particle-field systems in plasma physics. And it's also not restricted
for the space-time translation and rotation symmetries, but the general
symmetries. When the symmetries is developed, the conservation laws
for the systems can be established strictly.

This paper is organized as follows. In Sec.$\,$\ref{-----sec:General_Classical_PF_Sys_and_Weak_EL_eq},
we introduce the action of a general particle-field system. The weak
EL equation is developed as necessitated by the fact that classical
particles and fields live on different manifolds. Symmetries for the
system are discussed in Sec.$\,$\ref{----sec:Symmetries-and-conservation-laws},
and the links between conservation laws and symmetries are established.
Special symmetries and conservation laws for the KP and KD systems
are derived in Secs.$\,$\ref{-----sec:KP-systems} and \ref{-----sec:KD-systems}.

\section{General classical particle-field systems and weak Euler-Lagrange
equation \label{-----sec:General_Classical_PF_Sys_and_Weak_EL_eq}}

In general, the action of a classical particle-field system is
\begin{equation}
\mathcal{A}=\sum_{a}\int L_{a}\left(t,\boldsymbol{X}_{a}\left(t\right),\dot{\boldsymbol{X}}_{a}\left(t\right),\boldsymbol{\psi}\left(t,\boldsymbol{X}_{a}\left(t\right)\right)\right)dt+\int\mathcal{L}_{F}\left(t,\boldsymbol{x},\boldsymbol{\psi},\frac{\partial\boldsymbol{\psi}}{\partial t},\frac{\partial\boldsymbol{\psi}}{\partial\boldsymbol{x}}\right)dtd^{3}\boldsymbol{x},\label{Non-compact_action}
\end{equation}
where $\boldsymbol{X}_{a}\left(t\right)$ is the trajectory of the
$a$-th particle and $\boldsymbol{\psi}=\boldsymbol{\psi}\left(t,\boldsymbol{x}\right)$
is a field of scalar, vector, or tensor type. Apparently, the dynamics
of particles and fields are defined on different manifolds. The field
$\boldsymbol{\psi}$ is on the 4D space-time, whereas each particle's
trajectory is on the 1D time-axis. Thus, the integral of the Lagrangian
density $\mathcal{L}_{F}$ for the field $\boldsymbol{\psi}$ is over
space-time, and the integral of Lagrangian $L_{a}$ for the $a$-th
particle is over time only. 

Because of this fact, the action defined in Eq.~(\ref{Non-compact_action})
is not easily applicable to Noether's procedure of derving conservation
laws in space-time. To overcome this difficult, we multiply the first
part in the right-side of Eq.$\thinspace$(\ref{Non-compact_action})
by the identity

\begin{equation}
\int\delta_{a}d^{3}\boldsymbol{x}=1,\label{eq:2}
\end{equation}
where $\delta_{a}\equiv\delta\left(\boldsymbol{x}-\boldsymbol{X}_{a}\left(t\right)\right)$
is the Dirac $\delta$-function. The action $\mathcal{A}$ in Eq.$\thinspace$(\ref{Non-compact_action})
is then transformed into an integral over space-time, 
\begin{equation}
\mathcal{A}=\int\mathcal{L}\left(t,\boldsymbol{x},\boldsymbol{X}_{a}\left(t\right),\dot{\boldsymbol{X}}_{a}\left(t\right),\boldsymbol{\psi},\frac{\partial\boldsymbol{\psi}}{\partial t},\frac{\partial\boldsymbol{\psi}}{\partial\boldsymbol{x}}\right)dtd^{3}\boldsymbol{x},\label{Compact_Action}
\end{equation}
where $\mathcal{L}$ is the Lagrangian density of the particle-field
system defined as
\begin{align}
 & \mathcal{L}=\sum_{a}\mathcal{L}_{a}+\mathcal{L}_{F},\;\mathcal{L}_{a}=L_{a}\left(t,\boldsymbol{X}_{a}\left(t\right),\dot{\boldsymbol{X}}_{a}\left(t\right),\boldsymbol{\psi}\left(t,\boldsymbol{x}\right)\right)\delta_{a}.\label{Def.Ldensity_and_=00005CLa1}
\end{align}

Now we determine how the action given by Eq.$\thinspace$(\ref{Compact_Action})
varies in response to the variation of $\psi$,

\begin{equation}
\delta\mathcal{A}=\int\boldsymbol{E}_{\boldsymbol{\psi}}\left(\mathcal{L}\right)\cdot\delta\boldsymbol{\psi}dtd^{3}\boldsymbol{x},\label{eq:5}
\end{equation}
where the symbol ``$\cdot$'' stands for total contraction between
two tensors, and $\boldsymbol{E}_{\boldsymbol{\psi}}$ denotes the
Euler operator
\begin{equation}
\boldsymbol{E}_{\boldsymbol{\psi}}\left(\mathcal{L}\right)\equiv\frac{\partial\mathcal{L}}{\partial\boldsymbol{\psi}}-\frac{D}{Dt}\left[\frac{\partial\mathcal{L}}{\partial\left(\frac{\partial\boldsymbol{\psi}}{\partial t}\right)}\right]-\frac{D}{D\boldsymbol{x}}\cdot\left[\frac{\partial\mathcal{L}}{\partial\left(\frac{\partial\boldsymbol{\psi}}{\partial\boldsymbol{x}}\right)}\right].\label{Euler_Operator_field}
\end{equation}
Applying Hamilton's principle to Eq.$\thinspace$(\ref{eq:5}), we
immediately obtain the equation of motion for the field, 
\begin{equation}
\boldsymbol{E}_{\boldsymbol{\psi}}\left(\mathcal{L}\right)=0\,,\label{Field_EL_Eq.}
\end{equation}
by the arbitrariness of $\delta\boldsymbol{\psi}$.

Next, we derive the equation of motion for particles. There are two
ways to proceed. If we start from the action defined in Eq.~(\ref{Non-compact_action}),
the variation of $\mathcal{A}$ induced by $\delta\boldsymbol{X}_{a}$
is

\begin{equation}
\delta\mathcal{A}=\sum_{a}\int\left[\frac{\partial L_{a}}{\partial\boldsymbol{X}_{a}}-\frac{D}{Dt}\left(\frac{\partial L_{a}}{\partial\dot{\boldsymbol{X}}_{a}}\right)\right]\cdot\delta\boldsymbol{X}_{a}dt\,,\label{eq:8}
\end{equation}
and the EL equation of the $a$-th particle is

\begin{equation}
\frac{\partial L_{a}}{\partial\boldsymbol{X}_{a}}-\frac{D}{Dt}\left(\frac{\partial L_{a}}{\partial\dot{\boldsymbol{X}}_{a}}\right)=0\,.\label{EL_Eq._particles}
\end{equation}
Since Eq.$\thinspace$(\ref{EL_Eq._particles}) is not a differential
equation on space-time, it cannot be directly adopted in Noether's
procedure of deriving conservation laws.

The alternative way is to use the action defined in Eq.$\thinspace$(\ref{Compact_Action}),
which varies as 
\begin{equation}
\delta\mathcal{A}=\sum_{a}\int dt\delta\boldsymbol{X}_{a}\cdot\left[\int\left[\frac{\partial\mathcal{L}}{\partial\boldsymbol{X}_{a}}-\frac{D}{Dt}\left(\frac{\partial\mathcal{L}}{\partial\dot{\boldsymbol{X}}_{a}}\right)\right]d^{3}\boldsymbol{x}\right]\,,\label{eq:10}
\end{equation}
in response to the variation of $\boldsymbol{X}_{a}$. Here, the term
$\delta\boldsymbol{X}_{a}$ in Eq.$\thinspace$(\ref{eq:10}) was
moved outside from the integral $\int[\cdots]d^{3}\boldsymbol{x}$
because it is independent of $\boldsymbol{x}$. Hamilton's principle,
i.e., $\delta\mathcal{A}=0$ for the variation $\text{\ensuremath{\delta\boldsymbol{X}_{a}}}$,
requires the integral over the configuration space vanishes, 
\begin{equation}
\int\boldsymbol{E}_{\boldsymbol{X}_{a}}\left(\mathcal{L}\right)d^{3}\boldsymbol{x}=0.\label{Sub.EL_Eq.}
\end{equation}
Here, $\boldsymbol{E}_{\boldsymbol{X}_{a}}\left(\mathcal{L}\right)$
is the Euler operator with respect to $\boldsymbol{X}_{a}$, 
\begin{equation}
\boldsymbol{E}_{\boldsymbol{X}_{a}}\left(\mathcal{L}\right)\equiv\frac{\partial\mathcal{L}}{\partial\boldsymbol{X}_{a}}-\frac{D}{Dt}\left(\frac{\partial\mathcal{L}}{\partial\dot{\boldsymbol{X}}_{a}}\right).\label{Euler_operator_particles}
\end{equation}
Following Refs.~\citep{Qin2014,Fan2018}, Eq.~(\ref{Sub.EL_Eq.})
is called submanifold EL equation because it is defined only on the
time-axis after integrating over the spatial dimensions. Both Eqs.$\thinspace$(\ref{EL_Eq._particles})
and (\ref{Sub.EL_Eq.}) describe the equation of motion of the $a$-th
particle. The equivalence of the two equations can be easily proved
as follows,
\begin{eqnarray}
 &  & \frac{\partial L_{a}}{\partial\boldsymbol{X}_{a}}-\frac{D}{Dt}\left(\frac{\partial L_{a}}{\partial\dot{\boldsymbol{X}}_{a}}\right)=\frac{\partial}{\partial\boldsymbol{X}_{a}}\left(\int\mathcal{L}_{a}d^{3}\boldsymbol{x}\right)+\frac{D}{Dt}\left[\frac{\partial}{\partial\dot{\boldsymbol{X}}_{a}}\left(\int\mathcal{L}_{a}d^{3}\boldsymbol{x}\right)\right]\nonumber \\
 &  & =\int\left[\frac{\partial\mathcal{L}_{a}}{\partial\boldsymbol{X}_{a}}-\frac{D}{Dt}\left(\frac{\partial\mathcal{L}_{a}}{\partial\dot{\boldsymbol{X}}_{a}}\right)\right]d^{3}\boldsymbol{x}=\int\boldsymbol{E}_{\boldsymbol{X}_{a}}\left(\mathcal{L}\right)d^{3}\boldsymbol{x}.\label{eq:13}
\end{eqnarray}

In Eq.$\thinspace$(\ref{Sub.EL_Eq.}), the vanishing integral over
the configuration space suggests that the integrand $\boldsymbol{E}_{\boldsymbol{X}_{a}}\left(\mathcal{L}\right)$
could be a total divergence. We now derive such an explicit expression
for it. For the first term in $\boldsymbol{E}_{\boldsymbol{X}_{a}}\left(\mathcal{L}\right)$,
\begin{eqnarray}
\frac{\partial\mathcal{L}}{\partial\boldsymbol{X}_{a}} & = & \frac{\partial}{\partial\boldsymbol{X}_{a}}\left(L_{a}\delta_{a}\right)=L_{a}\frac{\partial\delta_{a}}{\partial\boldsymbol{X}_{a}}+\frac{\partial L_{a}}{\partial\boldsymbol{X}_{a}}\delta_{a}\nonumber \\
 & = & -L_{a}\frac{D\delta_{a}}{D\boldsymbol{x}}+\frac{\partial L_{a}}{\partial\boldsymbol{X}_{a}}\delta_{a}=\frac{D}{D\boldsymbol{x}}\cdot\left(-\mathcal{L}_{a}\boldsymbol{I}\right)+\frac{\partial L_{a}}{\partial\boldsymbol{X}_{a}}\delta_{a},\label{eq:14}
\end{eqnarray}
where $\boldsymbol{I}$ is the unit tensor and the identity $\partial\delta_{a}/\partial\boldsymbol{X}_{a}=-\partial\delta_{a}/\partial\boldsymbol{x}$
is used. For the second term in $\boldsymbol{E}_{\boldsymbol{X}_{a}}\left(\mathcal{L}\right)$,
\begin{eqnarray}
 &  & -\frac{D}{Dt}\left(\frac{\partial\mathcal{L}}{\partial\dot{\boldsymbol{X}}_{a}}\right)=-\frac{\partial L_{a}}{\partial\dot{\boldsymbol{X}}_{a}}\frac{D\delta_{a}}{Dt}-\frac{D}{Dt}\left(\frac{\partial L_{a}}{\partial\dot{\boldsymbol{X}}_{a}}\right)\delta_{a}\nonumber \\
 &  & =\frac{\partial L_{a}}{\partial\dot{\boldsymbol{X}}_{a}}\dot{\boldsymbol{X}}_{a}\cdot\frac{D\delta_{a}}{D\boldsymbol{x}}-\frac{D}{Dt}\left(\frac{\partial L_{a}}{\partial\dot{\boldsymbol{X}}_{a}}\right)\delta_{a}=\frac{D}{D\boldsymbol{x}}\cdot\left(\dot{\boldsymbol{X}}_{a}\frac{\partial\mathcal{L}_{a}}{\partial\dot{\boldsymbol{X}}_{a}}\right)-\frac{D}{Dt}\left(\frac{\partial L_{a}}{\partial\dot{\boldsymbol{X}}_{a}}\right)\delta_{a}.\label{eq:15}
\end{eqnarray}
Thus,
\begin{align}
\boldsymbol{E}_{\boldsymbol{X}_{a}}\left(\mathcal{L}\right) & =\frac{D}{D\boldsymbol{x}}\cdot\left(\dot{\boldsymbol{X}}_{a}\frac{\partial\mathcal{L}_{a}}{\partial\dot{\boldsymbol{X}}_{a}}-\mathcal{L}_{a}\boldsymbol{I}\right)+\left[\frac{\partial L_{a}}{\partial\boldsymbol{X}_{a}}-\frac{D}{Dt}\left(\frac{\partial L_{a}}{\partial\dot{\boldsymbol{X}}_{a}}\right)\right]\delta_{a}\nonumber \\
 & =\frac{D}{D\boldsymbol{x}}\cdot\left(\dot{\boldsymbol{X}}_{a}\frac{\partial\mathcal{L}_{a}}{\partial\dot{\boldsymbol{X}}_{a}}-\mathcal{L}_{a}\boldsymbol{I}\right),\label{weak_EL_Eq.}
\end{align}
As expected, the integrand is not zero but a total divergence. We
will refer Eq.$\thinspace$(\ref{weak_EL_Eq.}) as weak Euler-Lagrange
equation, which as a differential equation is equivalent to the submanifold
EL equation (\ref{Sub.EL_Eq.}). The qualifier ``weak'' indicates
that only the spatial integral of $\boldsymbol{E}_{\boldsymbol{X}_{a}}\left(\mathcal{L}\right)$
in Eq.$\thinspace$(\ref{Sub.EL_Eq.}) is zero \citep{Qin2014,Fan2018}.

The weak EL equation is indispensable in establishing the connections
between symmetries and local conservation laws in space-time for the
classical particle-field systems under investigation. Especially,
the non-vanishing right-hand-side of the weak EL equation induces
a new current in the corresponding conservation laws \citep{Qin2014,Fan2018}. 

\section{Symmetries and conservation laws for particle-field systems\label{----sec:Symmetries-and-conservation-laws}}

We now turn to the symmetries of the particle-field systems. A symmetry
of the action $\mathcal{A}\left[\boldsymbol{X}_{a},\boldsymbol{\psi}\right]$
is a group of transformation 
\begin{equation}
\left(t,\boldsymbol{x};\boldsymbol{X}_{a},\boldsymbol{\psi}\right)\longmapsto\left(\tilde{t},\tilde{\boldsymbol{x}};\tilde{\boldsymbol{X}}_{a},\tilde{\boldsymbol{\psi}}\right)\coloneqq g_{\epsilon}\cdot\left(t,\boldsymbol{x};\boldsymbol{X}_{a},\boldsymbol{\psi}\right),\label{eq:17}
\end{equation}
such that 
\begin{equation}
\int\mathcal{L}\left(t,\boldsymbol{x};\boldsymbol{X}_{a},\frac{d\boldsymbol{X}_{a}}{dt},\boldsymbol{\psi},\frac{\partial\boldsymbol{\psi}}{\partial t},\frac{\partial\boldsymbol{\psi}}{\partial\boldsymbol{x}}\right)dtd^{3}\boldsymbol{x}=\int\mathcal{L}\left(\tilde{t},\tilde{\boldsymbol{x}};\tilde{\boldsymbol{X}}_{a},\frac{d\tilde{\boldsymbol{X}}_{a}}{d\tilde{t}},\tilde{\boldsymbol{\psi}},\frac{\partial\tilde{\boldsymbol{\psi}}}{\partial\tilde{t}},\frac{\partial\tilde{\boldsymbol{\psi}}}{\partial\tilde{\boldsymbol{x}}}\right)d\tilde{t}d^{3}\tilde{\boldsymbol{x}}.\label{Symmetry_Condi.}
\end{equation}
Here $g_{\epsilon}$ constitutes a continuous group of transformations
parameterized by $\epsilon$ \citep{Olver1993}. To derive the corresponding
local conservation law, an infinitesimal symmetry criterion is needed.
We first define the infinitesimal generator induced by the group of
transformations as

\begin{equation}
\boldsymbol{v}\coloneqq\frac{d}{d\epsilon}|_{0}g_{\epsilon}\cdot\left(t,\boldsymbol{x};\boldsymbol{X}_{a},\boldsymbol{\psi}\right)=\xi^{t}\frac{\partial}{\partial t}+\boldsymbol{\kappa}\cdot\frac{\partial}{\partial\boldsymbol{x}}+\sum_{a}\boldsymbol{\theta}_{a}\cdot\frac{\partial}{\partial\boldsymbol{X}_{a}}+\boldsymbol{\phi}\cdot\frac{\partial}{\partial\boldsymbol{\psi}}.\label{eq:19}
\end{equation}
The symmetry condition (\ref{Symmetry_Condi.}) can be written as
\begin{equation}
\mathrm{pr}^{\left(1\right)}\boldsymbol{v}\left(\mathcal{L}\right)+\mathcal{L}\frac{D}{D\boldsymbol{\chi}}\cdot\boldsymbol{\xi}=0,\label{Inf.Symmetry_Criterion}
\end{equation}
where $\boldsymbol{\xi}$, $\boldsymbol{\chi}$ are 4D vectors in
space-time, i.e., $\xi^{\mu}=\left(\xi^{t},\boldsymbol{\kappa}\right)$
and $\chi^{\mu}=\left(t,\boldsymbol{x}\right)\thinspace(\mu=0,1,2,3)$
in a given coordinate system. Here, $\mathrm{pr}^{\left(1\right)}\boldsymbol{v}$,
as a vector field on the jet space, is the prolongation of the vector
field $\boldsymbol{v}$ on $\left\{ \left(t,\boldsymbol{x};\boldsymbol{X}_{a},\boldsymbol{\psi}\right)\right\} $,
\begin{equation}
\mathrm{pr}^{\left(1\right)}\boldsymbol{v}\coloneqq\frac{d}{d\epsilon}|_{0}\left(\tilde{t},\tilde{\boldsymbol{x}};\tilde{\boldsymbol{X}}_{a},\frac{d\tilde{\boldsymbol{X}}_{a}}{d\tilde{t}},\tilde{\boldsymbol{\psi}},\frac{\partial\tilde{\boldsymbol{\psi}}}{\partial\tilde{t}},\frac{\partial\tilde{\boldsymbol{\psi}}}{\partial\tilde{\boldsymbol{x}}}\right).\label{Prolon.Def.}
\end{equation}
The following expression for $\mathrm{pr}^{\left(1\right)}\boldsymbol{v}$
can be derived \citep{Olver1993},
\begin{equation}
\mathrm{pr}^{\left(1\right)}\boldsymbol{v}=\boldsymbol{v}+\sum_{a}\boldsymbol{\theta}_{a1}\cdot\frac{\partial}{\partial\dot{\boldsymbol{X}}_{a}}+\boldsymbol{\phi}_{\nu}\cdot\frac{\partial}{\partial\left(\frac{\partial\boldsymbol{\psi}}{\partial\chi^{\nu}}\right)},\label{eq:Prolon.Formu.}
\end{equation}
where $\boldsymbol{\theta}_{a1}$, and $\boldsymbol{\phi}_{\nu}$
are defined by 
\begin{equation}
\begin{aligned} & \boldsymbol{\theta}_{a1}=\xi^{t}\ddot{\boldsymbol{X}}_{a}+\dot{\boldsymbol{q}}_{a},\:\boldsymbol{\phi}_{\nu}=\xi^{\mu}\frac{D}{D\chi^{\nu}}\left(\frac{\partial\boldsymbol{\psi}}{\partial\chi^{\mu}}\right)+\frac{D\boldsymbol{Q}}{D\chi^{\nu}},\end{aligned}
\label{eq:23}
\end{equation}
and
\begin{equation}
\boldsymbol{q}_{a}=\boldsymbol{\theta}_{a}-\xi^{t}\dot{\boldsymbol{X}}_{a},\:\boldsymbol{Q}=\boldsymbol{\phi}-\xi^{\mu}\frac{\partial\boldsymbol{\psi}}{\partial\chi^{\mu}},\label{characteristics}
\end{equation}
are the corresponding characteristics of the vector field $\boldsymbol{v}$
\citep{Olver1993}.

Sometimes the system we encounter does not admit an given symmetry,
and the symmetry condition (\ref{Symmetry_Condi.}) is only valid
for part of the Lagrangian. That is, 

\begin{equation}
\mathcal{L}=\mathcal{L}_{\mathrm{S}}+\mathcal{F},\label{eq:25}
\end{equation}
where $\mathcal{L}_{S}$ is a part of the Lagrangian density satisfying

\begin{equation}
\int\mathcal{L}_{\mathrm{S}}\left(t,\boldsymbol{x};\boldsymbol{X}_{a},\frac{d\boldsymbol{X}_{a}}{dt},\boldsymbol{\psi},\frac{\partial\boldsymbol{\psi}}{\partial t},\frac{\partial\boldsymbol{\psi}}{\partial\boldsymbol{x}}\right)dtd^{3}\boldsymbol{x}=\int\mathcal{L}_{\mathrm{S}}\left(\tilde{t},\tilde{\boldsymbol{x}};\tilde{\boldsymbol{X}}_{a},\frac{d\tilde{\boldsymbol{X}}_{a}}{d\tilde{t}},\tilde{\boldsymbol{\psi}},\frac{\partial\tilde{\boldsymbol{\psi}}}{\partial\tilde{t}},\frac{\partial\tilde{\boldsymbol{\psi}}}{\partial\tilde{\boldsymbol{x}}}\right)d\tilde{t}d^{3}\tilde{\boldsymbol{x}}.\label{eq:26}
\end{equation}
In this situation, the infinitesimal symmetry criterion is

\begin{equation}
\mathrm{pr}^{\left(1\right)}\boldsymbol{v}\left(\mathcal{L}\right)+\mathcal{L}\frac{D}{D\boldsymbol{\chi}}\cdot\boldsymbol{\xi}=\mathrm{pr}^{\left(1\right)}\boldsymbol{v}\left(\mathcal{F}\right)+\mathcal{F}\frac{D}{D\boldsymbol{\chi}}\cdot\boldsymbol{\xi}.\label{Part.Inf.Symmetry_Criterion}
\end{equation}

Having derived the weak EL equation (\ref{weak_EL_Eq.}) and infinitesimal
symmetry criterion (\ref{Inf.Symmetry_Criterion}) or (\ref{Part.Inf.Symmetry_Criterion}),
we now establish the connection between symmetries and local conservation
laws. Substituting Eqs.$\thinspace$(\ref{eq:19}), (\ref{eq:Prolon.Formu.})
and (\ref{eq:23}) into the first term of Eq.$\thinspace$(\ref{Inf.Symmetry_Criterion}),
we have
\begin{align}
 & \mathrm{pr}^{\left(1\right)}\boldsymbol{v}\left(\mathcal{L}\right)=\xi^{\mu}\frac{\partial\mathcal{L}}{\partial\chi^{\mu}}+\sum_{a}\boldsymbol{\theta}_{a}\cdot\frac{\partial\mathcal{L}}{\partial\boldsymbol{X}_{a}}+\boldsymbol{\phi}\cdot\frac{\partial\mathcal{L}}{\partial\boldsymbol{\psi}}\nonumber \\
 & +\sum_{a}\left(\xi^{t}\ddot{\boldsymbol{X}}_{a}+\dot{\boldsymbol{q}}_{a}\right)\cdot\frac{\partial\mathcal{L}}{\partial\dot{\boldsymbol{X}}_{a}}+\left[\xi^{\mu}\frac{D}{D\chi^{\nu}}\left(\frac{\partial\boldsymbol{\psi}}{\partial\chi^{\mu}}\right)+\frac{D\boldsymbol{Q}}{D\chi^{\nu}}\right]\cdot\frac{\partial\mathcal{L}}{\partial\left(\frac{\partial\boldsymbol{\psi}}{\partial\chi^{\nu}}\right)}\nonumber \\
 & =\xi^{\mu}\frac{D\mathcal{L}}{D\chi^{\mu}}+\sum_{a}\left(\boldsymbol{\theta}_{a}-\xi^{t}\dot{\boldsymbol{X}}_{a}\right)\cdot\frac{\partial\mathcal{L}}{\partial\boldsymbol{X}_{a}}+\sum_{a}\dot{\boldsymbol{q}}_{a}\cdot\frac{\partial\mathcal{L}}{\partial\dot{\boldsymbol{X}}_{a}}+\left(\boldsymbol{\phi}-\xi^{\mu}\frac{\partial\boldsymbol{\psi}}{\partial\chi^{\mu}}\right)\cdot\frac{\partial\mathcal{L}}{\partial\boldsymbol{\psi}}+\frac{D\boldsymbol{Q}}{D\chi^{\nu}}\cdot\frac{\partial\mathcal{L}}{\partial\left(\frac{\partial\boldsymbol{\psi}}{\partial\chi^{\nu}}\right)}\nonumber \\
 & =\xi^{\mu}\frac{D\mathcal{L}}{D\chi^{\mu}}+\frac{D}{Dt}\left(\sum_{a}\boldsymbol{q}_{a}\cdot\frac{\partial\mathcal{L}}{\partial\dot{\boldsymbol{X}}_{a}}\right)+\frac{D}{D\chi^{\nu}}\left[\boldsymbol{Q}\cdot\frac{\partial\mathcal{L}}{\partial\left(\frac{\partial\boldsymbol{\psi}}{\partial\chi^{\nu}}\right)}\right]+\sum_{a}\boldsymbol{q}_{a}\cdot\boldsymbol{E}_{\boldsymbol{X}_{a}}\left(\mathcal{L}\right)+\boldsymbol{Q}\cdot\boldsymbol{E}_{\boldsymbol{\psi}}\left(\mathcal{L}\right)\nonumber \\
 & =\frac{D}{Dt}\left[\mathcal{L}\xi^{t}+\frac{\partial\mathcal{L}}{\partial\left(\frac{\partial\boldsymbol{\psi}}{\partial t}\right)}\cdot\boldsymbol{Q}+\sum_{a}\frac{\partial\mathcal{L}}{\partial\dot{\boldsymbol{X}}_{a}}\cdot\boldsymbol{q}_{a}\right]+\frac{D}{D\boldsymbol{x}}\cdot\left[\mathcal{L}\boldsymbol{\kappa}+\frac{\partial\mathcal{L}}{\partial\left(\frac{\partial\boldsymbol{\psi}}{\partial\boldsymbol{x}}\right)}\cdot\boldsymbol{Q}\right]\nonumber \\
 & +\sum_{a}\boldsymbol{q}_{a}\cdot\boldsymbol{E}_{\boldsymbol{X}_{a}}\left(\mathcal{L}\right)+\boldsymbol{Q}\cdot\boldsymbol{E}_{\boldsymbol{\psi}}\left(\mathcal{L}\right)-\mathcal{L}\frac{D}{D\boldsymbol{\chi}}\cdot\boldsymbol{\xi},\label{eq:28}
\end{align}
where Eq.$\thinspace$(\ref{characteristics}) is used for the third
step. Equation (\ref{Inf.Symmetry_Criterion}) now reads
\begin{align}
 & \frac{D}{Dt}\left[\mathcal{L}\xi^{t}+\frac{\partial\mathcal{L}}{\partial\left(\frac{\partial\boldsymbol{\psi}}{\partial t}\right)}\cdot\boldsymbol{Q}+\sum_{a}\frac{\partial\mathcal{L}}{\partial\dot{\boldsymbol{X}}_{a}}\cdot\boldsymbol{q}_{a}\right]\nonumber \\
 & +\frac{D}{D\boldsymbol{x}}\cdot\left[\mathcal{L}\boldsymbol{\kappa}+\frac{\partial\mathcal{L}}{\partial\left(\frac{\partial\boldsymbol{\psi}}{\partial\boldsymbol{x}}\right)}\cdot\boldsymbol{Q}\right]+\sum_{a}\boldsymbol{q}_{a}\cdot\boldsymbol{E}_{\boldsymbol{X}_{a}}\left(\mathcal{L}\right)+\boldsymbol{Q}\cdot\boldsymbol{E}_{\boldsymbol{\psi}}\left(\mathcal{L}\right)=0.\label{eq:29}
\end{align}
According to the EL equation (\ref{Field_EL_Eq.}) for $\boldsymbol{\psi}$,
the last term in Eq.$\thinspace$(\ref{eq:29}) vanishes. However,
due to the weak EL equation (\ref{weak_EL_Eq.}), the third term in
Eq.$\thinspace$(\ref{eq:29}) is not zero. If the characteristics
$\boldsymbol{q}_{a}$ is independent of \textbf{$\boldsymbol{x}$}
and $\boldsymbol{\psi}$, this term can be written as a divergence
form, i.e., 
\begin{equation}
\boldsymbol{q}_{a}\cdot\boldsymbol{E}_{\boldsymbol{X}_{a}}\left(\mathcal{L}\right)=\frac{D}{D\boldsymbol{x}}\cdot\left[\left(\dot{\boldsymbol{X}}_{a}\frac{\partial\mathcal{L}_{a}}{\partial\dot{\boldsymbol{X}}_{a}}-\mathcal{L}_{a}\boldsymbol{I}\right)\cdot\boldsymbol{q}_{a}\right]\,,\label{eq:30}
\end{equation}
which induces a new current absent in the standard field theory. Substituting
Eq.$\thinspace$(\ref{eq:30}) into Eq.$\thinspace$(\ref{eq:29}),
we finally arrive at the conservation law

\begin{align}
 & \frac{D}{Dt}\left[\mathcal{L}\xi^{t}+\frac{\partial\mathcal{L}}{\partial\left(\frac{\partial\boldsymbol{\psi}}{\partial t}\right)}\cdot\boldsymbol{Q}+\sum_{a}\frac{\partial\mathcal{L}}{\partial\dot{\boldsymbol{X}}_{a}}\cdot\boldsymbol{q}_{a}\right]\nonumber \\
 & +\frac{D}{D\boldsymbol{x}}\cdot\left[\mathcal{L}\boldsymbol{\kappa}+\frac{\partial\mathcal{L}}{\partial\left(\frac{\partial\boldsymbol{\psi}}{\partial\boldsymbol{x}}\right)}\cdot\boldsymbol{Q}+\sum_{a}\left(\dot{\boldsymbol{X}}_{a}\frac{\partial\mathcal{L}_{a}}{\partial\dot{\boldsymbol{X}}_{a}}-\mathcal{L}_{a}\boldsymbol{I}\right)\cdot\boldsymbol{q}_{a}\right]=0.\label{eq:Conservation}
\end{align}
If the symmetry condition of the system is Eq.$\thinspace$(\ref{Part.Inf.Symmetry_Criterion})
instead, the corresponding conservation law of the system should be
changed to 
\begin{align}
 & \frac{D}{Dt}\left[\mathcal{L}\xi^{t}+\frac{\partial\mathcal{L}}{\partial\left(\frac{\partial\boldsymbol{\psi}}{\partial t}\right)}\cdot\boldsymbol{Q}+\sum_{a}\frac{\partial\mathcal{L}}{\partial\dot{\boldsymbol{X}}_{a}}\cdot\boldsymbol{q}_{a}\right]\nonumber \\
 & +\frac{D}{D\boldsymbol{x}}\cdot\left[\mathcal{L}\boldsymbol{\kappa}+\frac{\partial\mathcal{L}}{\partial\left(\frac{\partial\boldsymbol{\psi}}{\partial\boldsymbol{x}}\right)}\cdot\boldsymbol{Q}+\sum_{a}\left(\dot{\boldsymbol{X}}_{a}\frac{\partial\mathcal{L}_{a}}{\partial\dot{\boldsymbol{X}}_{a}}-\mathcal{L}_{a}\boldsymbol{I}\right)\cdot\boldsymbol{q}_{a}\right]=\mathrm{pr}^{\left(1\right)}\boldsymbol{v}\left(\mathcal{F}\right)+\mathcal{F}\frac{D}{D\boldsymbol{\text{\ensuremath{\chi}}}}\cdot\boldsymbol{\xi}\,,\label{eq:Part.Conservation}
\end{align}
which states that the space-time divergence of the flux equals the
input form the source.

\section{Symmetries and conservation laws for Klimontovich-Poisson system
\label{-----sec:KP-systems}}

The Klimontovich-Poisson system, as a reduced system of the Klimontovich-Maxwell
system, has been applied extensively in plasma physics. The local
energy-momentum conservation laws for the KP system has important
implications. The action and Lagrangian density of the KP system are
given by 
\begin{equation}
\begin{aligned} & \mathcal{A}=\int\mathcal{L}_{\mathrm{KP}}dtd^{3}\boldsymbol{x},\thinspace\mathcal{L}_{KP}=\sum_{a}\mathcal{L}_{a}+\mathcal{L}_{\mathrm{F}},\\
 & \mathcal{L}_{a}=\left[\frac{1}{2}m_{a}\dot{\boldsymbol{X}}_{a}^{2}+\frac{q_{a}}{c}\dot{\boldsymbol{X}}_{a}\cdot\boldsymbol{A}_{0}\left(\boldsymbol{x}\right)-q_{a}\varphi\right]\delta_{a},\:\mathcal{L}_{F}=\frac{\left(\boldsymbol{\nabla}\varphi\right)^{2}}{8\pi},
\end{aligned}
\label{eq:KP_Lagrangian}
\end{equation}
where $\boldsymbol{A}_{0}$ is the vector potential for a given external
magnetic field $\boldsymbol{B}_{0}=\boldsymbol{\nabla}\times\boldsymbol{A}_{0}$,
and the field $\boldsymbol{\psi}$ in this case is the scalar potential
$\varphi$.

As a benchmark against the result in Ref.$\thinspace$\citep{Qin2014},
we first discuss the time translation symmetry and the energy conservation
law for the KP system. Substituting Eq.$\thinspace$(\ref{eq:KP_Lagrangian})
into Eq.$\thinspace$(\ref{weak_EL_Eq.}), we immediately obtain the
weak EL equation,
\begin{align}
 & \boldsymbol{E}_{\boldsymbol{X}_{a}}\left(\mathcal{L}_{\mathrm{KP}}\right)=\frac{\partial\mathcal{L}_{\mathrm{KP}}}{\partial\boldsymbol{X}_{a}}-\frac{d}{dt}\frac{\partial\mathcal{L}_{\mathrm{KP}}}{\partial\dot{\boldsymbol{X}}_{a}}\nonumber \\
 & =\frac{D}{D\boldsymbol{x}}\cdot\left[\dot{\boldsymbol{X}}_{a}\left(m_{a}\dot{\boldsymbol{X}}_{a}+\frac{q_{a}}{c}\boldsymbol{A}_{0}\left(\boldsymbol{x}\right)\right)\delta_{a}-\left(\frac{1}{2}m_{a}\dot{\boldsymbol{X}}_{a}^{2}+\frac{q_{a}}{c}\dot{\boldsymbol{X}}_{a}\cdot\boldsymbol{A}_{0}\left(\boldsymbol{x}\right)-q_{a}\varphi\right)\delta_{a}\boldsymbol{I}\right],
\end{align}
which is the same as the result in Ref.~\citep{Qin2014}. It is also
straightforward to verify that the action of the KP system is invariant
under the time translation, 
\begin{equation}
\left(t,\boldsymbol{x};\boldsymbol{X}_{a},\varphi\right)\mapsto\left(\tilde{t},\tilde{\boldsymbol{x}};\tilde{\boldsymbol{X}}_{a},\tilde{\varphi}\right)=g_{\epsilon}\cdot\left(t,\boldsymbol{x};\boldsymbol{X}_{a},\varphi\right)=\left(t+\epsilon,\boldsymbol{x};\boldsymbol{X}_{a},\varphi\right),\:\epsilon\in\mathbb{R}\,.\label{eq:tme_translation}
\end{equation}
The infinitesimal generator of the group transformation is 
\begin{equation}
\boldsymbol{v}=\frac{\partial}{\partial t}\,,\label{eq:36}
\end{equation}
whose prolongation in the jet space is 
\begin{equation}
\mathrm{pr}^{\left(1\right)}\boldsymbol{v}=\frac{\partial}{\partial t}\,.\label{eq:37}
\end{equation}
 The infinitesimal criterion (\ref{Inf.Symmetry_Criterion}) of the
symmetry, naturally satisfied by the Lagrangian, is 
\begin{equation}
\frac{\partial\mathcal{L}}{\partial t}=0.\label{eq:38}
\end{equation}
The characteristic $\boldsymbol{q}_{a}=\boldsymbol{\theta}_{a}-\xi^{t}\dot{\boldsymbol{X}}_{a}=-\dot{\boldsymbol{X}}_{a}$
is independent of $\boldsymbol{x}$ and $\varphi$. Substituting Eqs.$\thinspace$(\ref{eq:KP_Lagrangian})
and (\ref{eq:36}) into Eq.$\thinspace$(\ref{eq:Conservation}),
we obtain the energy conservation law,
\begin{align}
 & \frac{D}{Dt}\left[\frac{\left(\boldsymbol{\nabla}\varphi\right)^{2}}{8\pi}-\sum_{a}\left(\frac{1}{2}m_{a}\dot{\boldsymbol{X}}_{a}^{2}+q_{a}\varphi\right)\delta_{a}\right]+\frac{D}{D\boldsymbol{x}}\cdot\left[-\frac{\boldsymbol{\nabla}\varphi}{4\pi}\varphi_{,t}-\sum_{a}\left(\frac{1}{2}m_{a}\dot{\boldsymbol{X}}_{a}^{2}+q_{a}\varphi\right)\delta_{a}\dot{\boldsymbol{X}}_{a}\right]=0,\label{eq:39}
\end{align}
where $\varphi_{,t}\equiv\partial_{t}\varphi$. Subtracting the identity
\begin{equation}
\frac{D}{Dt}\left\{ \frac{D}{D\boldsymbol{x}}\cdot\left[\varphi\frac{\partial\mathcal{L}}{\partial\left(\nabla\varphi\right)}\right]\right\} +\frac{D}{D\boldsymbol{x}}\cdot\left\{ -\frac{D}{Dt}\left[\varphi\frac{\partial\mathcal{L}}{\partial\left(\nabla\varphi\right)}\right]\right\} =0\label{eq:40}
\end{equation}
from Eq.$\thinspace$(\ref{eq:39}), the energy conservation is (equivalently)
\begin{align}
 & \frac{D}{Dt}\left[\sum_{a}\frac{1}{2}m_{a}\dot{\boldsymbol{X}}_{a}^{2}\delta_{a}+\frac{\left(\boldsymbol{\nabla}\varphi\right)^{2}}{8\pi}\right]+\frac{D}{D\boldsymbol{x}}\cdot\left[\sum_{a}\left(\frac{1}{2}m_{a}\dot{\boldsymbol{X}}_{a}^{2}+q_{a}\varphi\right)\delta_{a}\dot{\boldsymbol{X}}_{a}-\frac{1}{4\pi}\varphi\boldsymbol{\nabla}\varphi_{,t}\right]=0\,.\label{eq:41}
\end{align}
It can be easily seen that the result of Eq.(\ref{eq:41}) agrees
with the result given in Ref.$\thinspace$\citep{Qin2014}, which
evidently support the general theory we established here (see Eqs.(\ref{weak_EL_Eq.})
and (\ref{eq:Conservation})).

We now discuss the connection between the rotational symmetry and
the angular momentum conservation law of the KP system, which has
not been studied previously. The Lagrangian density is first split
into two parts,
\begin{equation}
\begin{aligned} & \mathcal{L}_{\mathrm{KP}}=\mathcal{L}_{\mathrm{S}}+\mathcal{F},\\
 & \mathcal{L}_{\mathrm{S}}=\sum_{a}\left[\frac{1}{2}m_{a}\dot{\boldsymbol{X}}_{a}^{2}-q_{a}\varphi\right]\delta_{a}+\frac{\left(\boldsymbol{\nabla}\varphi\right)^{2}}{8\pi},\mathcal{F}=\sum_{a}\frac{q_{a}}{c}\dot{\boldsymbol{X}}_{a}\cdot\boldsymbol{A}_{0}\left(\boldsymbol{x}\right)\delta_{a},
\end{aligned}
\label{eq:42}
\end{equation}
where $\mathcal{L}_{S}$ is invariant under the rotational transformation
and the symmetry is responsible for the conservation of local angular
momentum. However, the term $\mathcal{F}$ does comply with the rotational
symmetry, and it represents a torque due to the external magnetic
field generating input of angular momentum to the system. We now choose
a global Cartesian coordinate to describe the rotation. In this coordinate
system, all vectors, such as $\boldsymbol{x},\boldsymbol{X}_{a},\text{ and }\boldsymbol{A}_{0},$
are represented by $1\times3$ matrices. The rotational transformations
of the system is defined by 
\begin{equation}
\left(t,\boldsymbol{x};\boldsymbol{X}_{a},\varphi\right)\mapsto\left(\tilde{t},\tilde{\boldsymbol{x}};\tilde{\boldsymbol{X}}_{a},\tilde{\varphi}\right)=g_{\epsilon}\cdot\left(t,\boldsymbol{x};\boldsymbol{X}_{a},\varphi\right)=\left(t,\boldsymbol{R}_{\epsilon}\cdot\boldsymbol{x};\boldsymbol{R}_{\epsilon}\cdot\boldsymbol{X}_{a},\varphi\right),\:\epsilon\in\mathbb{R},\label{eq:43}
\end{equation}
where $\boldsymbol{R}_{\epsilon}$ is a continuous one parameter subgroup
of $\mathrm{SO}\left(3\right)$, the rotational group in the 3D Euclidean
space. At $\epsilon=0,$$\boldsymbol{R}_{0}=\boldsymbol{I}$ is the
identity matrix. Substituting Eq.$\thinspace$(\ref{eq:43}) into
Eq.$\thinspace$(\ref{eq:Prolon.Formu.}), the infinitesimal generator
and its prolongation are 
\begin{equation}
\boldsymbol{v}=\left(\boldsymbol{\Omega}\cdot\boldsymbol{x}\right)\cdot\frac{\partial}{\partial\boldsymbol{x}}+\sum_{a}\left(\boldsymbol{\Omega}\cdot\boldsymbol{X}_{a}\right)\cdot\frac{\partial}{\partial\boldsymbol{X}_{a}},\label{eq:44}
\end{equation}
and 
\begin{equation}
\mathrm{pr}^{\left(1\right)}\boldsymbol{v}=\left(\boldsymbol{\Omega}\cdot\boldsymbol{x}\right)\cdot\frac{\partial}{\partial\boldsymbol{x}}+\sum_{a}\left(\boldsymbol{\Omega}\cdot\boldsymbol{X}_{a}\right)\cdot\frac{\partial}{\partial\boldsymbol{X}_{a}}+\sum_{a}\left(\boldsymbol{\Omega}\cdot\dot{\boldsymbol{X}}_{a}\right)\cdot\frac{\partial}{\partial\dot{\boldsymbol{X}}_{a}}+\left(\boldsymbol{\Omega}\cdot\boldsymbol{\nabla}\varphi\right)\cdot\frac{\partial}{\partial\boldsymbol{\nabla}\varphi},\label{eq:45}
\end{equation}
where 
\begin{equation}
\boldsymbol{\Omega}=\frac{d}{d\epsilon}|_{0}\boldsymbol{R}_{\epsilon}\label{eq:46}
\end{equation}
is a $3\times3$ anti-symmetric matrix, i.e., an element in the Lie
algebra $so(3)$. The characteristic $\boldsymbol{q}_{a}\equiv\boldsymbol{\theta}_{a}-\xi^{t}\dot{\boldsymbol{X}}_{a}=\boldsymbol{\Omega}\cdot\boldsymbol{X}_{a}$
is independent of $\boldsymbol{x}$ and $\varphi$. Substituting Eqs.$\thinspace$(\ref{eq:42})
and (\ref{eq:45}) into the left-hand side of Eq.$\thinspace$(\ref{Inf.Symmetry_Criterion}),
we have
\begin{align}
 & \mathrm{pr}^{\left(1\right)}\boldsymbol{v}\left(\mathcal{L}_{\mathrm{KP}}\right)+\mathcal{L}_{\mathrm{KP}}\frac{D}{D\boldsymbol{\chi}}\cdot\boldsymbol{\xi}\nonumber \\
 & =-\boldsymbol{\Omega}:\left[\boldsymbol{x}\frac{\partial\mathcal{L}_{\mathrm{KP}}}{\partial\boldsymbol{x}}+\sum_{a}\boldsymbol{X}_{a}\frac{\partial\mathcal{L}_{\mathrm{KP}}}{\partial\boldsymbol{X}_{a}}+\sum_{a}\dot{\boldsymbol{X}}_{a}\frac{\partial\mathcal{L}_{\mathrm{KP}}}{\partial\dot{\boldsymbol{X}}_{a}}+\boldsymbol{\nabla}\varphi\frac{\partial\mathcal{L}_{\mathrm{KP}}}{\partial\boldsymbol{\nabla}\varphi}\right]\nonumber \\
 & =-\boldsymbol{\Omega}:\left[\boldsymbol{x}\frac{\partial\mathcal{L}_{S}}{\partial\boldsymbol{x}}+\sum_{a}\boldsymbol{X}_{a}\frac{\partial\mathcal{L}_{S}}{\partial\boldsymbol{X}_{a}}+\sum_{a}\dot{\boldsymbol{X}}_{a}\frac{\partial\mathcal{L}_{S}}{\partial\dot{\boldsymbol{X}}_{a}}+\boldsymbol{\nabla}\varphi\frac{\partial\mathcal{L}_{S}}{\partial\boldsymbol{\nabla}\varphi}\right]+\mathrm{pr}^{\left(1\right)}\boldsymbol{v}\left(\mathcal{F}\right)+\mathcal{F}\frac{D}{D\boldsymbol{\chi}}\cdot\boldsymbol{\xi}\nonumber \\
 & =\mathrm{pr}^{\left(1\right)}\boldsymbol{v}\left(\mathcal{F}\right)+\mathcal{F}\frac{D}{D\boldsymbol{\chi}}\cdot\boldsymbol{\xi}-\boldsymbol{\Omega}:\left[\sum_{a}m_{a}\dot{\boldsymbol{X}}_{a}\dot{\boldsymbol{X}}_{a}\delta\left(\boldsymbol{x}-\boldsymbol{X}_{a}\right)+\frac{\boldsymbol{\nabla}\varphi\boldsymbol{\nabla}\varphi}{4\pi}\right]\nonumber \\
 & -\sum_{a}\left(\frac{1}{2}m_{a}\dot{\boldsymbol{X}}_{a}^{2}-q_{a}\varphi\right)\boldsymbol{\Omega}:\left[\boldsymbol{x}\frac{\partial\delta_{a}}{\partial\boldsymbol{x}}+\boldsymbol{X}_{a}\frac{\partial\delta_{a}}{\partial\boldsymbol{X}_{a}}\right],\label{eq:47}
\end{align}
where operator``$:$'' between two matrices is defined to be
\begin{equation}
\boldsymbol{C}:\boldsymbol{D}=\mathrm{tr}\left(\boldsymbol{C}\cdot\boldsymbol{D}^{T}\right)\,.
\end{equation}
The third term of the right-hand side of Eq.$\thinspace$(\ref{eq:47})
is zero because $\boldsymbol{\Omega}:\boldsymbol{H}=0$ for any symmetric
matrix $\boldsymbol{H}$. The last term of Eq.$\thinspace$(\ref{eq:47})
also vanishes, 
\begin{align}
 & \boldsymbol{\Omega}:\left[\boldsymbol{x}\frac{\partial\delta_{a}}{\partial\boldsymbol{x}}+\boldsymbol{X}_{a}\frac{\partial\delta_{a}}{\partial\boldsymbol{X}_{a}}\right]=\frac{d}{d\theta}|_{0}\delta\left(\boldsymbol{R}_{\theta}\cdot\boldsymbol{x}-\boldsymbol{R}_{\theta}\cdot\boldsymbol{X}_{a}\right)=\frac{d}{d\theta}|_{0}\frac{\delta\left(\boldsymbol{x}-\boldsymbol{X}_{a}\right)}{\mathrm{det}\boldsymbol{R}_{\epsilon}}=0.\label{eq:49}
\end{align}
Equation$\thinspace$(\ref{eq:47}) then reduces to
\begin{equation}
\mathrm{pr}^{\left(1\right)}\boldsymbol{v}\left(\mathcal{L}_{\mathrm{KP}}\right)+\mathcal{L}_{\mathrm{KP}}\frac{D}{D\boldsymbol{x}}\cdot\boldsymbol{\xi}=\mathrm{pr}^{\left(1\right)}\boldsymbol{v}\left(\mathcal{F}\right)+\mathcal{F}\frac{D}{D\boldsymbol{x}}\cdot\boldsymbol{\xi},\label{eq:50}
\end{equation}
which is in the form of Eq.$\thinspace$(\ref{Part.Inf.Symmetry_Criterion}).
Therefore, the corresponding conservation law assumes the form of
Eq.$\thinspace$(\ref{eq:Part.Conservation}). For the rotational
symmetry under investigation, the right-hand side of Eq.$\thinspace$(\ref{eq:50})
can be transformed into 
\begin{eqnarray}
 &  & \mathrm{pr}^{\left(1\right)}\boldsymbol{v}\left(\mathcal{F}\right)+\mathcal{F}\frac{D}{D\boldsymbol{x}}\cdot\boldsymbol{\xi}=\mathrm{pr}^{\left(1\right)}\boldsymbol{v}\left(\mathcal{F}\right)=-\boldsymbol{\Omega}:\left[\boldsymbol{x}\frac{\partial\mathcal{F}}{\partial\boldsymbol{x}}+\sum_{a}\boldsymbol{X}_{a}\frac{\partial\mathcal{F}}{\partial\boldsymbol{X}_{a}}+\sum_{a}\dot{\boldsymbol{X}}_{a}\frac{\partial\mathcal{F}}{\partial\dot{\boldsymbol{X}}_{a}}+\boldsymbol{\nabla}\varphi\frac{\partial\mathcal{F}}{\partial\boldsymbol{\nabla}\varphi}\right]\nonumber \\
 &  & =\boldsymbol{\Omega}:\sum_{a}\frac{q_{a}}{c}\left[\dot{\boldsymbol{X}}_{a}\cdot\left(\boldsymbol{\nabla}\boldsymbol{A}_{0}\right)^{T}\boldsymbol{x}+\boldsymbol{A}_{0}\left(\boldsymbol{x}\right)\dot{\boldsymbol{X}}_{a}\right]\delta_{a}-\sum_{a}\left[\frac{q_{a}}{c}\dot{\boldsymbol{X}}_{a}\cdot\boldsymbol{A}_{0}\left(\boldsymbol{x}\right)\right]\boldsymbol{\Omega}:\left[\boldsymbol{x}\frac{\partial\delta_{a}}{\partial\boldsymbol{x}}+\boldsymbol{X}_{a}\frac{\partial\delta_{a}}{\partial\boldsymbol{X}_{sp}}\right]\nonumber \\
 &  & =\boldsymbol{\Omega}:\sum_{a}\frac{q_{a}}{c}\left[\dot{\boldsymbol{X}}_{a}\cdot\left(\boldsymbol{\nabla}\boldsymbol{A}_{0}\right)^{T}\boldsymbol{x}+\boldsymbol{A}_{0}\left(\boldsymbol{x}\right)\dot{\boldsymbol{X}}_{a}\right]\delta_{2},\label{eq:51}
\end{eqnarray}
where we used Eq.$\thinspace$(\ref{eq:49}). The conservation law
is 
\begin{align}
 & \frac{D}{Dt}\left\{ \boldsymbol{\Omega}:\left[\left(\sum_{a}m_{a}\dot{\boldsymbol{X}}_{a}\delta_{2}\right)\boldsymbol{x}\right]\right\} +\frac{D}{D\boldsymbol{x}}\cdot\left\{ \left[\left(\sum_{a}m_{a}\dot{\boldsymbol{X}}_{a}\dot{\boldsymbol{X}}_{a}\delta_{a}\right)\boldsymbol{x}+\frac{\left(\boldsymbol{\nabla}\varphi\right)^{2}}{8\pi}\boldsymbol{I}-\frac{\boldsymbol{\nabla}\varphi\boldsymbol{\nabla}\varphi}{4\pi}\right]\boldsymbol{x}:\boldsymbol{\Omega}\right\} \nonumber \\
 & +\frac{D}{Dt}\left\{ \boldsymbol{\Omega}:\left[\left(\sum_{a}\frac{q_{a}}{c}\boldsymbol{A}_{0}\left(\boldsymbol{x}\right)\delta_{a}\right)\boldsymbol{x}\right]\right\} +\frac{D}{D\boldsymbol{x}}\cdot\left\{ \left[\left(\sum_{a}\frac{q_{a}}{c}\dot{\boldsymbol{X}}_{a}\boldsymbol{A}_{0}\left(\boldsymbol{x}\right)\delta_{a}\right)\boldsymbol{x}:\boldsymbol{\Omega}\right]\right\} \nonumber \\
 & =\boldsymbol{\Omega}:\sum_{a}\frac{q_{a}}{c}\left[\dot{\boldsymbol{X}}_{a}\cdot\left(\boldsymbol{\nabla}\boldsymbol{A}_{0}\right)^{T}\boldsymbol{x}+\boldsymbol{A}_{0}\left(\boldsymbol{x}\right)\dot{\boldsymbol{X}}_{a}\right]\delta_{a}.\label{eq:52}
\end{align}
The last two terms on the left-hand side of Eq.$\thinspace$(\ref{eq:52})
can be combined,
\begin{align}
 & \frac{D}{Dt}\left\{ \boldsymbol{\Omega}:\left[\left(\sum_{a}\frac{q_{a}}{c}\boldsymbol{A}_{0}\left(\boldsymbol{x}\right)\right)\boldsymbol{x}\right]\right\} +\frac{D}{D\boldsymbol{x}}\cdot\left\{ \left[\left(\sum_{a}\frac{q_{a}}{c}\dot{\boldsymbol{X}}_{a}\boldsymbol{A}_{0}\left(\boldsymbol{x}\right)\delta_{a}\right)\boldsymbol{x}\right]:\boldsymbol{\Omega}\right\} \nonumber \\
 & =\sum_{a}\frac{q_{a}}{c}\left[\dot{\boldsymbol{X}}_{a}\cdot\boldsymbol{\nabla}\boldsymbol{A}_{0}\left(\boldsymbol{x}\right)\boldsymbol{x}\delta_{a}+\boldsymbol{A}_{0}\left(\boldsymbol{x}\right)\dot{\boldsymbol{X}}_{a}\delta_{a}\right]:\boldsymbol{\Omega}\,,\label{eq:53}
\end{align}
and the conservation law is simplified into
\begin{align}
 & \frac{D}{Dt}\left\{ \boldsymbol{\Omega}:\left[\left(\sum_{a}m_{a}\dot{\boldsymbol{X}}_{a}\delta_{a}\right)\boldsymbol{x}\right]\right\} +\frac{D}{D\boldsymbol{x}}\cdot\left\{ \left[\left(\sum_{a}m_{a}\dot{\boldsymbol{X}}_{a}\dot{\boldsymbol{X}}_{a}\delta_{a}\right)\boldsymbol{x}+\frac{\left(\boldsymbol{\nabla}\varphi\right)^{2}}{8\pi}\boldsymbol{I}-\frac{\boldsymbol{\nabla}\varphi\boldsymbol{\nabla}\varphi}{4\pi}\right]\boldsymbol{x}:\boldsymbol{\Omega}\right\} \nonumber \\
 & =\boldsymbol{\Omega}:\sum_{a}\frac{q_{a}}{c}\left\{ \dot{\boldsymbol{X}}_{a}\cdot\left[\left(\boldsymbol{\nabla}\boldsymbol{A}_{0}\right)^{T}-\left(\boldsymbol{\nabla}\boldsymbol{A}_{0}\right)\right]\boldsymbol{x}\right\} \delta_{a}.\label{eq:54}
\end{align}
Equation (\ref{eq:52}) can be equivalently written as 
\begin{align}
 & \boldsymbol{\omega}\cdot\left\{ \frac{D}{Dt}\left[\boldsymbol{x}\times\left(\sum_{a}m_{a}\dot{\boldsymbol{X}}_{a}\delta_{a}\right)\right]-\frac{D}{D\boldsymbol{x}}\cdot\left\{ \left[\sum_{a}m_{a}\dot{\boldsymbol{X}}_{a}\dot{\boldsymbol{X}}_{a}\delta_{a}+\frac{\left(\boldsymbol{\nabla}\varphi\right)^{2}}{8\pi}\boldsymbol{I}-\frac{\boldsymbol{\nabla}\varphi\boldsymbol{\nabla}\varphi}{4\pi}\right]\times\boldsymbol{x}\right\} \right\} \nonumber \\
 & =\boldsymbol{\omega}\cdot\sum_{a}\frac{q_{s}}{c}\boldsymbol{x}\times\left\{ \dot{\boldsymbol{X}}_{a}\cdot\left[\left(\boldsymbol{\nabla}\boldsymbol{A}_{0}\right)^{T}-\left(\boldsymbol{\nabla}\boldsymbol{A}_{0}\right)\right]\right\} \delta_{a},\label{eq:55}
\end{align}
where the vector $\boldsymbol{\omega}$ is defined as
\begin{equation}
\omega_{k}\equiv-\frac{1}{2}\sum_{i,j}\Omega_{ij}\epsilon_{ijk}\,.\label{eq:56}
\end{equation}
Here, $\epsilon_{ijk}$ is the Levi-Civita symbol. Equation (\ref{eq:56})
implies 
\begin{equation}
\Omega_{ij}=-\sum_{k}\omega_{k}\epsilon_{ijk}.\label{eq:57}
\end{equation}
In Eq.$\thinspace$(\ref{eq:55}), the cross operator ``$\times$''
is defined by
\begin{equation}
\left(\boldsymbol{a}\times\boldsymbol{b}\right)_{i}=\sum_{j,k}\epsilon_{ijk}a_{j}b_{k},\:\left(\boldsymbol{C}\times\boldsymbol{a}\right)_{ij}=\sum_{k,l}\epsilon_{ikl}C_{jk}a_{l},\:i,j,k,l=1,2,3
\end{equation}
for any 3-vectors $\boldsymbol{a}$,$\boldsymbol{b}$ and $3\times3$
matrix\textbf{ $\boldsymbol{C}$}. Due to the arbitrariness of the
vector $\boldsymbol{\omega}$, Eq.~(\ref{eq:55}) implies 
\begin{align}
 & \frac{D}{Dt}\left(\boldsymbol{x}\times\boldsymbol{g}_{\mathrm{KP}}\right)+\frac{D}{D\boldsymbol{x}}\cdot\left(-\boldsymbol{T}_{\mathrm{KP}}\times\boldsymbol{x}\right)=\sum_{a}\frac{q_{a}}{c}\boldsymbol{x}\times\left[\dot{\boldsymbol{X}}_{a}\times\left(\boldsymbol{\nabla}\times\boldsymbol{A}_{0}\right)\right]\delta_{a},\label{eq:59}
\end{align}
where used is made of the following identity
\begin{equation}
\dot{\boldsymbol{X}}_{a}\cdot\left[\left(\boldsymbol{\nabla}\boldsymbol{A}_{0}\right)^{T}-\left(\boldsymbol{\nabla}\boldsymbol{A}_{0}\right)\right]=\dot{\boldsymbol{X}}_{a}\times\left(\boldsymbol{\nabla}\times\boldsymbol{A}_{0}\right)\,.
\end{equation}
The momentum density \textbf{$\boldsymbol{g}_{KP}$} and stress matrix
$\boldsymbol{T}_{KP}$ of the KP system in Eq.$\thinspace$(\ref{eq:59})
are defined as
\begin{equation}
\begin{aligned} & \boldsymbol{g}_{\mathrm{KP}}=\sum_{a}m_{a}\dot{\boldsymbol{X}}_{a}\delta_{a},\:\boldsymbol{T}_{\mathrm{KP}}=\sum_{a}m_{a}\dot{\boldsymbol{X}}_{a}\dot{\boldsymbol{X}}_{a}\delta_{a}+\frac{\left(\boldsymbol{\nabla}\varphi\right)^{2}}{8\pi}\boldsymbol{I}-\frac{\boldsymbol{\nabla}\varphi\boldsymbol{\nabla}\varphi}{4\pi}.\end{aligned}
\end{equation}

\section{Rotational symmetry and angular momentum conservation law for Klimontovich-Darwin
system\label{-----sec:KD-systems}}

Another well-known reduced model is the Klimontovich-Darwin (KD) system
\citep{Kaufman1971,Jackson1999,Qin2001,Krause2007}. For the KD system,
the action and Lagrangian density are given by
\begin{equation}
\begin{aligned} & \mathcal{A}=\int\mathcal{L}_{\mathrm{KD}}dtd^{3}\boldsymbol{x},\mathcal{L}_{\mathrm{KD}}=\sum_{a}\mathcal{L}_{a}+\mathcal{L}_{F},\\
 & \mathcal{L}_{a}=\left(\frac{1}{2}m_{a}\dot{\boldsymbol{X}}_{a}^{2}-q_{a}\varphi+\frac{q_{a}}{c}\dot{\boldsymbol{X}}_{a}\cdot\boldsymbol{A}\right)\delta_{a},\:\mathcal{L}_{\mathrm{F}}=\frac{1}{8\pi}\left[\left(\boldsymbol{\nabla}\varphi\right)^{2}+\frac{2}{c}\boldsymbol{\nabla}\varphi\cdot\partial_{t}\boldsymbol{A}-\left(\boldsymbol{\nabla}\times\boldsymbol{A}\right)^{2}\right].
\end{aligned}
\label{eq:KD_Lagrangian}
\end{equation}
In this case, the field $\boldsymbol{\psi}=\left(\varphi,\boldsymbol{A}\right)$
is the 4-potential. The vector potential $\boldsymbol{A}\left(t,\boldsymbol{x}\right)$
is part of the dynamics, which is different from the external field
$\boldsymbol{A}_{0}\left(t,\boldsymbol{x}\right)$ of the KP system
in Sec.$\thinspace$\ref{-----sec:KP-systems}. The rotational transformations
of the KD system is 
\begin{equation}
\left(\tilde{t},\tilde{\boldsymbol{x}};\tilde{\boldsymbol{X}}_{a},\tilde{\varphi},\tilde{\mathbf{A}}\right)=g_{\epsilon}\cdot\left(t,\boldsymbol{x};\boldsymbol{X}_{a},\varphi,\mathbf{A}\right)=\left(t,\boldsymbol{R}_{\epsilon}\cdot\boldsymbol{x};\boldsymbol{R}_{\epsilon}\cdot\boldsymbol{X}_{a},\varphi,\boldsymbol{R}_{\epsilon}\cdot\mathbf{A}\right),\:\epsilon\in\mathbb{R},\label{eq:63}
\end{equation}
where the definition of $\boldsymbol{R}_{\epsilon}$ is same as that
in Sec.$\thinspace$\ref{-----sec:KP-systems}. Note that the symmetry
transformation includes the rotation of the vector potential $\boldsymbol{A}\left(t,\boldsymbol{x}\right)$.
The infinitesimal generator and its prolongation (\ref{eq:63}) are
\begin{align}
 & \boldsymbol{v}=\left(\boldsymbol{\Omega}\cdot\boldsymbol{x}\right)\cdot\frac{\partial}{\partial\boldsymbol{x}}+\sum_{a}\left(\boldsymbol{\Omega}\cdot\boldsymbol{X}_{a}\right)\cdot\frac{\partial}{\partial\boldsymbol{X}_{a}}+\left(\boldsymbol{\Omega}\cdot\boldsymbol{A}\right)\cdot\frac{\partial}{\partial\boldsymbol{A}},\label{eq:64}\\
 & \mathrm{pr}^{\left(1\right)}\boldsymbol{v}=\boldsymbol{v}+\sum_{a}\left(\boldsymbol{\Omega}_{a}\cdot\dot{\boldsymbol{X}}_{a}\right)\cdot\frac{\partial}{\partial\dot{\boldsymbol{X}}_{a}}+\left(\boldsymbol{\Omega}\cdot\boldsymbol{\nabla}\varphi\right)\cdot\frac{\partial}{\partial\left(\boldsymbol{\nabla}\varphi\right)}+\left(\boldsymbol{\Omega}\cdot\partial_{t}\boldsymbol{A}\right)\cdot\frac{\partial}{\partial\left(\partial_{t}\boldsymbol{A}\right)}\nonumber \\
 & +\left[\boldsymbol{\Omega}\cdot\left(\boldsymbol{\nabla}\boldsymbol{A}\right)-\left(\boldsymbol{\nabla}\boldsymbol{A}\right)\cdot\boldsymbol{\Omega}\right]:\frac{\partial}{\partial\left(\boldsymbol{\nabla}\boldsymbol{A}\right)}\,.\label{eq:65}
\end{align}
The characteristic $\boldsymbol{q}_{a}=\boldsymbol{\theta}_{a}-\xi^{t}\dot{\boldsymbol{X}}_{a}=\boldsymbol{\Omega}\cdot\boldsymbol{X}_{a}$
is independent of $\boldsymbol{x}$, $\varphi$ and \textbf{$\boldsymbol{A}$}.
Substituting Eqs.$\thinspace$(\ref{eq:KD_Lagrangian}), (\ref{eq:64})
and (\ref{eq:65}) into the left-hand side of Eq.$\thinspace$(\ref{Inf.Symmetry_Criterion}),
we have
\begin{eqnarray}
 &  & \mathrm{pr}^{\left(1\right)}\boldsymbol{v}\left(\mathcal{L}_{\mathrm{KD}}\right)+\mathcal{L}_{\mathrm{KD}}\frac{D}{D\boldsymbol{\chi}}\cdot\boldsymbol{\xi}\nonumber \\
 &  & =-\boldsymbol{\Omega}:\left\{ \boldsymbol{x}\frac{\partial\mathcal{L}_{\mathrm{KD}}}{\partial\boldsymbol{x}}+\sum_{a}\boldsymbol{X}_{a}\frac{\partial\mathcal{L}_{\mathrm{KD}}}{\partial\boldsymbol{X}_{a}}+\boldsymbol{A}\frac{\partial\mathcal{L}_{\mathrm{KD}}}{\partial\boldsymbol{A}}+\sum_{a}\dot{\boldsymbol{X}}_{a}\frac{\partial\mathcal{L}_{\mathrm{KD}}}{\partial\dot{\boldsymbol{X}}_{sp}}+\boldsymbol{\nabla}\varphi\frac{\partial\mathcal{L}_{\mathrm{KD}}}{\partial\left(\boldsymbol{\nabla}\varphi\right)}\right.\nonumber \\
 &  & =-\boldsymbol{\Omega}:\left\{ \sum_{a}\left(\frac{1}{2}m_{a}\dot{\boldsymbol{X}}_{a}^{2}-q_{a}\varphi+\frac{q_{a}}{c}\dot{\boldsymbol{X}}_{a}\cdot\boldsymbol{A}\right)\left(\boldsymbol{x}\frac{\partial\delta_{a}}{\partial\boldsymbol{x}}+\boldsymbol{X}_{a}\frac{\partial\delta_{a}}{\partial\boldsymbol{X}_{a}}\right)\right.\nonumber \\
 &  & \vphantom{\frac{\partial\mathcal{L}_{KM}}{\partial\left(\boldsymbol{\nabla}\boldsymbol{A}\right)}}+\sum_{a}\left[\frac{q_{a}}{c}\left(\boldsymbol{A}\dot{\boldsymbol{X}}_{a}+\dot{\boldsymbol{X}}_{a}\boldsymbol{A}\right)\delta_{a}+m_{a}\dot{\boldsymbol{X}}_{a}\dot{\boldsymbol{X}}_{a}\delta_{a}\right]\vphantom{\frac{\partial\mathcal{L}_{KM}}{\partial\left(\boldsymbol{\nabla}\boldsymbol{A}\right)}}+\frac{1}{4\pi}\left[\boldsymbol{\nabla}\varphi\boldsymbol{\nabla}\varphi+\frac{1}{c}\left(\boldsymbol{\nabla}\varphi\partial_{t}\boldsymbol{A}+\partial_{t}\boldsymbol{A}\boldsymbol{\nabla}\varphi\right)\right]\nonumber \\
 &  & \left.\vphantom{\frac{\partial\mathcal{L}_{KM}}{\partial\left(\boldsymbol{\nabla}\boldsymbol{A}\right)}}+\frac{1}{4\pi}\left[\left(\boldsymbol{\nabla}\boldsymbol{A}\right)-\left(\boldsymbol{\nabla}\boldsymbol{A}\right)^{T}\right]\cdot\left[\left(\boldsymbol{\nabla}\boldsymbol{A}\right)-\left(\boldsymbol{\nabla}\boldsymbol{A}\right)^{T}\right]\right\} ,\label{eq:66}
\end{eqnarray}
where used is made of the following equations
\begin{equation}
\begin{aligned} & \frac{\partial\mathcal{L}_{\mathrm{KD}}}{\partial\left(\boldsymbol{\nabla}\boldsymbol{A}\right)}=-\frac{1}{4\pi}\left(\boldsymbol{\epsilon}:\boldsymbol{\nabla}\boldsymbol{A}\right)\cdot\boldsymbol{\epsilon}=-\frac{1}{4\pi}\boldsymbol{\epsilon}\cdot\left(\boldsymbol{\epsilon}:\boldsymbol{\nabla}\boldsymbol{A}\right),\:\boldsymbol{\epsilon}\cdot\left(\boldsymbol{\epsilon}:\boldsymbol{\nabla}\boldsymbol{A}\right)=\left(\boldsymbol{\nabla}\boldsymbol{A}\right)-\left(\boldsymbol{\nabla}\boldsymbol{A}\right)^{T}.\end{aligned}
\label{eq:67}
\end{equation}
The first term on the right-hand side of Eq.$\thinspace$(\ref{eq:66})
is zero because of Eq.$\thinspace$(\ref{eq:49}). The last three
terms on the right-hand side of Eq.$\thinspace$(\ref{eq:66}) also
vanish because they are the traces of matrix products between a symmetric
and an anti-symmetric matrices. The vanishing right-hand-side of Eq.$\,$(\ref{eq:66})
verifies that Eq.~(\ref{eq:63}) is a symmetry of the system. Substituting
$\boldsymbol{\xi},$ $\boldsymbol{\theta}_{a}$ and $\boldsymbol{\phi}$
in Eq.$\thinspace$(\ref{eq:64}) into Eq.$\thinspace$(\ref{eq:Conservation}),
we obtain the angular momentum conservation law of the rotational
symmetry for the KD system,
\begin{align}
 & \frac{D}{Dt}\left\{ -\boldsymbol{\Omega}:\boldsymbol{x}\left[\sum_{a}\left(m_{a}\dot{\boldsymbol{X}}_{a}+\frac{q_{a}}{c}\boldsymbol{A}\right)\delta_{a}-\frac{1}{4\pi c}\boldsymbol{\nabla}\boldsymbol{A}\cdot\boldsymbol{\nabla}\varphi\right]-\frac{1}{4\pi c}\boldsymbol{\Omega}:\left(\boldsymbol{A}\boldsymbol{\nabla}\varphi\right)\right\} \nonumber \\
 & +\frac{D}{D\boldsymbol{x}}\cdot\left\{ \left[\frac{1}{8\pi}\left[\left(\boldsymbol{\nabla}\varphi\right)^{2}+\frac{2}{c}\boldsymbol{\nabla}\varphi\cdot\partial_{t}\boldsymbol{A}-\left(\boldsymbol{\nabla}\times\boldsymbol{A}\right)^{2}\right]\boldsymbol{I}+\sum_{a}\left(m_{a}\dot{\boldsymbol{X}}_{a}\dot{\boldsymbol{X}}_{a}+\frac{q_{a}}{c}\dot{\boldsymbol{X}}_{a}\boldsymbol{A}\right)\delta_{a}\right.\right.\nonumber \\
 & \left.\vphantom{\frac{1}{8\pi}}\left.-\frac{1}{4\pi}\boldsymbol{\epsilon}:\left[\left(\boldsymbol{\epsilon}:\boldsymbol{\nabla}\boldsymbol{A}\right)\left(\boldsymbol{\nabla}\boldsymbol{A}\right)^{T}\right]-\frac{1}{4\pi}\left(\boldsymbol{\nabla}\varphi+\frac{1}{c}\partial_{t}\boldsymbol{A}\right)\boldsymbol{\nabla}\varphi\right]\boldsymbol{x}:\boldsymbol{\Omega}-\frac{1}{4\pi}\left[\boldsymbol{\epsilon}\cdot\left(\boldsymbol{\epsilon}:\boldsymbol{\nabla}\boldsymbol{A}\right)\boldsymbol{A}\right]:\boldsymbol{\Omega}\right\} =0.\label{eq:68}
\end{align}

To put the conservation law into a symmetric form, we add the following
identity
\begin{align}
\frac{D}{Dt}\left\{ \frac{D}{D\boldsymbol{x}}\cdot\left[\frac{\partial\mathcal{L}_{KD}}{\partial\boldsymbol{A}_{,t}}\boldsymbol{A}\boldsymbol{x}\right]\right\} :\boldsymbol{\Omega}+\frac{D}{D\boldsymbol{x}}\cdot\left\{ -\frac{D}{Dt}\left[\frac{\partial\mathcal{L}_{KD}}{\partial\boldsymbol{A}_{,t}}\boldsymbol{A}\boldsymbol{x}\right]\right\} :\boldsymbol{\Omega}=0 & \,,\label{eq:69}
\end{align}
to Eq.$\thinspace$(\ref{eq:68}) to get 
\begin{align}
 & \frac{D}{Dt}\left\{ -\boldsymbol{\Omega}:\boldsymbol{x}\left[\sum_{a}m_{a}\dot{\boldsymbol{X}}_{a}\delta_{a}+\frac{\left(-\boldsymbol{\nabla}\varphi\right)\times\left(\boldsymbol{\nabla}\times\boldsymbol{A}\right)}{4\pi c}-\frac{1}{4\pi c}\boldsymbol{A}\frac{D}{D\boldsymbol{x}}\cdot\left(\frac{1}{c}\partial_{t}\boldsymbol{A}\right)\right]\right\} \nonumber \\
 & +\frac{D}{D\boldsymbol{x}}\cdot\left\{ \left[\frac{1}{8\pi}\left[\left(\boldsymbol{\nabla}\varphi\right)^{2}+\frac{2}{c}\boldsymbol{\nabla}\varphi\cdot\boldsymbol{A}_{,t}+\left(\boldsymbol{\nabla}\times\boldsymbol{A}\right)\right]\boldsymbol{I}-\frac{1}{4\pi}\left(-\boldsymbol{\nabla}\varphi-\frac{1}{c}\boldsymbol{A}_{,t}\right)\left(-\boldsymbol{\nabla}\varphi-\frac{1}{c}\boldsymbol{A}_{,t}\right)\right.\right.\nonumber \\
 & \left.\vphantom{\frac{1}{8\pi}}\left.-\frac{1}{4\pi}\left(\boldsymbol{\nabla}\times\boldsymbol{A}\right)\left(\boldsymbol{\nabla}\times\boldsymbol{A}\right)+\sum_{a}m_{a}\dot{\boldsymbol{X}}_{a}\dot{\boldsymbol{X}}_{a}\delta_{a}+\frac{1}{4\pi c^{2}}\boldsymbol{A}_{,t}\boldsymbol{A}_{,t}\right]\boldsymbol{x}:\boldsymbol{\Omega}\right\} .\label{eq:70}
\end{align}
The detailed calculation of this symmetrization process is given in
Appendix \ref{sec:symmetrization_process}.

Using the relations between $\boldsymbol{\Omega}$ and $\boldsymbol{\omega}$
and the arbitrariness of $\boldsymbol{\omega}$, the angular momentum
conservation law (\ref{eq:70}) of the KD system can be equivalently
rewritten as
\begin{equation}
\frac{D}{Dt}\left(\boldsymbol{x}\times\boldsymbol{g}_{\mathrm{KD}}\right)+\frac{D}{D\boldsymbol{x}}\cdot\left(-\boldsymbol{T}_{\mathrm{KD}}\times\boldsymbol{x}\right)=0,\label{eq:71}
\end{equation}
where the momentum density $\boldsymbol{g}_{\mathrm{KD}}$ and the
stress matrix $\boldsymbol{T}_{KD}$ are defined by
\begin{equation}
\begin{aligned} & \boldsymbol{g}_{\mathrm{KD}}=\sum_{a}m_{a}\dot{\boldsymbol{X}}_{a}\delta_{a}+\frac{\left(-\boldsymbol{\nabla}\varphi\right)\times\boldsymbol{B}}{4\pi c}-\frac{1}{4\pi c^{2}}\boldsymbol{A}\partial_{t}\left(\boldsymbol{\nabla}\cdot\boldsymbol{A}\right),\\
 & \boldsymbol{T}_{KD}=\sum_{a}m_{a}\dot{\boldsymbol{X}}_{a}\dot{\boldsymbol{X}}_{a}\delta_{a}+\frac{\left(\boldsymbol{\nabla}\varphi\right)^{2}+2\boldsymbol{\nabla}\varphi\cdot\boldsymbol{A}_{,t}/c+\boldsymbol{B}^{2}}{8\pi}\boldsymbol{I}-\frac{\boldsymbol{EE}+\boldsymbol{BB}-\boldsymbol{A}_{,t}\boldsymbol{A}_{,t}/c^{2}}{4\pi}\,.
\end{aligned}
\label{eq:72}
\end{equation}
In obtaining Eq.~(\ref{eq:72}), the following identities were used,
\begin{equation}
\begin{aligned} & \left(\boldsymbol{\nabla}\boldsymbol{A}-\left(\boldsymbol{\nabla}\boldsymbol{A}\right)^{T}\right)\cdot\left(-\boldsymbol{\nabla}\varphi-\frac{1}{c}\partial_{t}\boldsymbol{A}\right)=\left(-\boldsymbol{\nabla}\varphi-\frac{1}{c}\partial_{t}\boldsymbol{A}\right)\times\left(\boldsymbol{\nabla}\times\boldsymbol{A}\right),\\
 & \boldsymbol{\epsilon}:\left[\left(\boldsymbol{\epsilon}:\boldsymbol{\nabla}\boldsymbol{A}\right)\left(\left(\boldsymbol{\nabla}\boldsymbol{A}\right)-\left(\boldsymbol{\nabla}\boldsymbol{A}\right)^{T}\right)\right]=\left(\boldsymbol{\nabla}\times\boldsymbol{A}\right)^{2}\boldsymbol{I}-\left(\boldsymbol{\nabla}\times\boldsymbol{A}\right)\left(\boldsymbol{\nabla}\times\boldsymbol{A}\right).
\end{aligned}
\label{eq:73}
\end{equation}

The angular momentum conservation law can be also derived directly
by using $\boldsymbol{x}$ to cross every terms in the momentum conservation
law of the KD system,

\begin{equation}
\frac{D}{Dt}\boldsymbol{g}_{\mathrm{KD}}+\frac{D}{D\boldsymbol{x}}\cdot\boldsymbol{T}_{\mathrm{KD}}=0.\label{eq:74}
\end{equation}
Using the general theory derived in see Sec.$\thinspace$\ref{----sec:Symmetries-and-conservation-laws},
we can prove that Eq.(\ref{eq:74}) is due to the space translation
symmetry

\begin{equation}
\left(\tilde{t},\tilde{\boldsymbol{x}};\tilde{\boldsymbol{X}}_{a},\tilde{\varphi},\tilde{\mathbf{A}}\right)=g_{\epsilon}\cdot\left(t,\boldsymbol{x};\boldsymbol{X}_{a},\varphi,\mathbf{A}\right)=\left(t,\boldsymbol{x}+\epsilon\boldsymbol{X}_{0};\boldsymbol{X}_{a}+\epsilon\boldsymbol{X}_{0},\varphi,\boldsymbol{R}_{\epsilon}\cdot\mathbf{A}\right)\label{eq:75}
\end{equation}
of the action for the Darwin's model (see Eq.$\thinspace$(\ref{eq:KD_Lagrangian})),
where $\epsilon\in\mathbb{R}$ and $\boldsymbol{X}_{0}$ is an arbitrary
constant vector. For this group of transformations, the corresponding
infinitesimal generator and its characteristics are calculated as

\begin{equation}
\begin{aligned} & \boldsymbol{v}=\boldsymbol{X}_{0}\cdot\frac{\partial}{\partial\boldsymbol{x}}+\boldsymbol{X}_{0}\cdot\sum_{a}\frac{\partial}{\partial\boldsymbol{X}_{0}},\\
 & \boldsymbol{q}_{a}=\boldsymbol{\theta}_{a}-\xi^{t}\dot{\boldsymbol{X}}_{a}=\boldsymbol{X}_{0},\\
 & \boldsymbol{Q}=\boldsymbol{\phi}-\xi^{\mu}\frac{\partial\boldsymbol{\psi}}{\partial\chi^{\mu}}=-\left(\boldsymbol{X}_{0}\cdot\boldsymbol{\nabla}\varphi,\boldsymbol{X}_{0}\cdot\boldsymbol{\nabla}\boldsymbol{A}\right).
\end{aligned}
\label{eq:76}
\end{equation}
Substituting Eq.$\thinspace$(\ref{eq:76}) into Eq.$\thinspace$(\ref{eq:Conservation}),
we get

\begin{alignat}{1}
 & \frac{D}{Dt}\left[-\frac{1}{4\pi c}\boldsymbol{\nabla}\varphi\cdot\left(\boldsymbol{\nabla}\boldsymbol{A}\right)^{T}+\sum_{a}\left(m_{a}\dot{\boldsymbol{X}}_{a}+\frac{q_{a}}{c}\boldsymbol{A}\right)\delta_{a}\right]+\frac{D}{D\boldsymbol{x}}\cdot\left[\mathcal{L}_{F}\boldsymbol{I}-\frac{1}{4\pi}\left(\boldsymbol{\nabla}\varphi+\frac{1}{c}\boldsymbol{A}_{,t}\right)\boldsymbol{\nabla}\varphi\right.\nonumber \\
 & \vphantom{\frac{1}{8\pi}}\left.-\frac{1}{4\pi}\boldsymbol{\epsilon}:\left[\left(\boldsymbol{\epsilon}:\boldsymbol{\nabla}\boldsymbol{A}\right)\left(\boldsymbol{\nabla}\boldsymbol{A}\right)^{T}\right]+\sum_{a}\left(m_{a}\dot{\boldsymbol{X}}_{a}\dot{\boldsymbol{X}}_{a}+\frac{q_{a}}{c}\dot{\boldsymbol{X}}_{a}\boldsymbol{A}\right)\delta_{a}\right].\label{eq:77}
\end{alignat}
At this point, Eq.$\thinspace$(\ref{eq:77}) is still different from
Eq.$\thinspace$(\ref{eq:74}). To get the momentum conservation law
in the form of Eq.$\thinspace$(\ref{eq:74}), an additional identity

\begin{equation}
\frac{D}{Dt}\left\{ \frac{D}{D\boldsymbol{x}}\cdot\left[\frac{\partial\mathcal{L}_{KD}}{\partial\boldsymbol{A}_{,t}}\boldsymbol{A}\right]\right\} +\frac{D}{D\boldsymbol{x}}\cdot\left\{ -\frac{D}{Dt}\left[\frac{\partial\mathcal{L}_{KD}}{\partial\boldsymbol{A}_{,t}}\boldsymbol{A}\right]\right\} =0\label{eq:78}
\end{equation}
needs to be added into Eq.(\ref{eq:77}).

Note that the momentum equation Eq.(\ref{eq:77}) is different from
that of Ref.$\thinspace$\citep{Kaufman1971}. To illustrate this
difference, we may add another identity

\begin{equation}
\frac{D}{Dt}\left\{ \frac{1}{4\pi c}\frac{D}{D\boldsymbol{x}}\cdot\left[\left(\boldsymbol{A}\cdot\boldsymbol{\nabla}\varphi\right)\boldsymbol{I}-\boldsymbol{A}\boldsymbol{\nabla}\varphi\right]\right\} +\frac{D}{D\boldsymbol{x}}\cdot\left\{ -\frac{1}{4\pi c}\frac{D}{Dt}\left[\left(\boldsymbol{A}\cdot\boldsymbol{\nabla}\varphi\right)\boldsymbol{I}-\boldsymbol{A}\boldsymbol{\nabla}\varphi\right]\right\} =0\label{eq:79}
\end{equation}
to Eq.$\thinspace$(\ref{eq:77}) to get a different form of Eq.$\thinspace$(\ref{eq:77})
as
\begin{equation}
\frac{D}{Dt}\boldsymbol{g}_{\mathrm{KD}}^{'}+\frac{D}{D\boldsymbol{x}}\cdot\boldsymbol{T}_{\mathrm{KD}}^{'}=0.\label{eq:80}
\end{equation}
Here, the momentum density $\boldsymbol{g}_{KD}^{'}$ and the stress
matrix $\boldsymbol{T}_{KD}^{'}$ are defined by
\begin{equation}
\begin{aligned}\boldsymbol{g}_{\mathrm{KD}}^{'}= & \sum_{a}\left(m_{a}\dot{\boldsymbol{X}}_{a}+\frac{q_{a}}{c}\boldsymbol{A}\right)\delta_{a}-\frac{1}{4\pi c}\left(\boldsymbol{\nabla}\cdot\boldsymbol{A}\right)\boldsymbol{\nabla}\varphi,\\
\boldsymbol{T}_{\mathrm{KD}}^{'}= & \sum_{a}m_{a}\dot{\boldsymbol{X}}_{a}\dot{\boldsymbol{X}}_{a}\delta_{a}+\frac{1}{8\pi}\left[\left(\boldsymbol{\nabla}\varphi\right)^{2}-\frac{2}{c}\boldsymbol{A}\cdot\partial_{t}\left(\boldsymbol{\nabla}\varphi\right)+\boldsymbol{B}^{2}\right]\boldsymbol{I}\\
 & -\frac{1}{4\pi}\left[\boldsymbol{\nabla}\varphi\boldsymbol{\nabla}\varphi+\boldsymbol{BB}-\frac{1}{c}\boldsymbol{A}\partial_{t}\left(\boldsymbol{\nabla}\varphi\right)-\frac{1}{c}\partial_{t}\left(\boldsymbol{\nabla}\varphi\right)\boldsymbol{A}\right].
\end{aligned}
\label{eq:81}
\end{equation}
In this process, we used

\begin{equation}
\frac{\partial\mathcal{L_{\mathrm{KD}}}}{\partial\boldsymbol{A}}\boldsymbol{A}=\sum_{a}\frac{q_{a}}{c}\delta_{a}\dot{\boldsymbol{X}}_{a}\boldsymbol{A}
\end{equation}
and the EL equation for $\boldsymbol{A}.$ In Ref.$\thinspace$\citep{Kaufman1971},
the authors chose the Coulomb gauge, i.e., $\boldsymbol{\nabla}\cdot\boldsymbol{A}=0$
for Eq.$\thinspace$(\ref{eq:81}), to get the momentum density 
\begin{equation}
\boldsymbol{g}_{\mathrm{KD}}^{'}=\sum_{a}\left(m_{a}\dot{\boldsymbol{X}}_{a}+\frac{q_{a}}{c}\boldsymbol{A}\right)\delta_{a}.\text{}\label{eq:82}
\end{equation}
However, Darwin's model is gauge dependent, i.e., the action (\ref{eq:KD_Lagrangian})
is not invariant under the gauge transformation

\begin{equation}
\boldsymbol{A}\rightarrow\boldsymbol{A}+\boldsymbol{\nabla}f,\varphi\rightarrow\varphi-\frac{\partial f}{\partial t},\label{eq:83}
\end{equation}
which implies that we can not choose a gauge condition for the system.

\section{Conclusions\label{sec:Conclusions}}

In this study, we developed a general field theory for classical particle-field
systems, and established the connections between general symmetries
and local conservation laws in space-time for the systems. Compared
with the standard classical field theory, the distinguish feature
of the classical particle-field systems is that the particles and
fields reside on different manifolds. The fields are defined on the
4D space-time, whereas each particle's trajectory is defined on the
1D time-axis. As a consequence, the standard Noether's procedure for
deriving local conservation laws from symmetries do not apply straightforwardly
without modification. To overcome this difficulty, a weak Euler-Lagrange
equation for particles is developed on the 4D space-time, which plays
a pivotal role in establishing the connections between symmetries
and local conservation laws in space-time. Especially, the non-vanishing
Euler derivative in the weak EL equation generates a new current in
the corresponding conservation laws. 

Several examples from plasma physics are studied as special cases
of the general field theory. As a benchmark, the time translation
symmetry of the Klimontovich-Poisson (KP) system and the corresponding
local energy conservation law were obtained by the general theory
and compared with the results in Ref.$\thinspace$\citep{Qin2014}.
As new applications, the relations between the rotational symmetry
and angular momentum conservation for the KP system and Klimontovich-Darwin
(KD) system are established. For the KP system, the conservation law
is manifested as the balance between space-time divergence of the
angular momentum flux and the input due to the torque of the external
magnetic field. For the KD system, it is found that the rotational
symmetry admitted by the system needs to include the rotation of the
vector potential. Such a rotation is a representation of the rotational
group in the fiber of the vector bundle at each space-time location.

It is also worth to mention that the exact conservation laws (such
as energy and momentum conservation laws) in driftkinetic theory and
gyrokinetic theory can also be derived from this theory systematically.
This topic will be discussed in the future work.
\begin{acknowledgments}
This research is supported by National Natural Science Foundation
of China (NSFC-11775219 and NSFC-11575186), the U.S. Department of
Energy (DE-AC02-09CH11466), and National Key Research and Development
Program (2016YFA0400600, 2016YFA0400601 and 2016YFA0400602).
\end{acknowledgments}

\appendix

\section{the symmetrization process of Eq.$\thinspace$(\ref{eq:68}) \label{sec:symmetrization_process}}

In this appendix, we give a detailed derivation of Eq.$\thinspace$(\ref{eq:70}).
The first term of Eq.$\thinspace$(\ref{eq:69}) is
\begin{eqnarray}
 &  & \frac{D}{Dt}\left\{ \frac{D}{D\boldsymbol{x}}\cdot\left[\frac{\partial\mathcal{L}_{KD}}{\partial\boldsymbol{A}_{,t}}\boldsymbol{A}\boldsymbol{x}\right]\right\} :\boldsymbol{\Omega}=\boldsymbol{\Omega}:\frac{D}{Dt}\left\{ \frac{D}{D\boldsymbol{x}}\cdot\left[\frac{\partial\mathcal{L}_{KD}}{\partial\boldsymbol{A}_{,t}}\right]\boldsymbol{A}\boldsymbol{x}+\frac{\partial\mathcal{L}_{KD}}{\partial\boldsymbol{A}_{,t}}\cdot\boldsymbol{\nabla}\boldsymbol{A}\boldsymbol{x}+\boldsymbol{A}\frac{\partial\mathcal{L}_{KD}}{\partial\boldsymbol{A}_{,t}}\right\} \nonumber \\
 &  & =\boldsymbol{\Omega}:\frac{D}{Dt}\left\{ \frac{1}{c}\frac{D}{D\boldsymbol{x}}\cdot\left(\frac{\partial\mathcal{L}_{KD}}{\partial\left(\boldsymbol{\nabla}\varphi\right)}-\frac{1}{4\pi c}\partial_{t}\boldsymbol{A}\right)\boldsymbol{A}\boldsymbol{x}+\frac{\partial\mathcal{L}_{KD}}{\partial\boldsymbol{A}_{,t}}\cdot\boldsymbol{\nabla}\boldsymbol{A}\boldsymbol{x}+\boldsymbol{A}\frac{\partial\mathcal{L}_{KD}}{\partial\left(\boldsymbol{A}_{,t}\right)}\right\} \nonumber \\
 &  & =\boldsymbol{\Omega}:\frac{D}{Dt}\left\{ \frac{1}{c}\frac{\partial\mathcal{L}_{KD}}{\partial\varphi}\boldsymbol{A}\boldsymbol{x}-\frac{1}{4\pi c}\frac{D}{D\boldsymbol{x}}\cdot\left(\frac{1}{c}\partial_{t}\boldsymbol{A}\right)\boldsymbol{A}\boldsymbol{x}+\frac{1}{4\pi c}\boldsymbol{\nabla}\varphi\cdot\boldsymbol{\nabla}\boldsymbol{A}\boldsymbol{x}+\frac{1}{4\pi c}\boldsymbol{A}\boldsymbol{\nabla}\varphi\right\} \nonumber \\
 &  & =\frac{D}{Dt}\left\{ -\boldsymbol{\Omega}:\boldsymbol{x}\left[-\sum_{a}\frac{q_{a}}{c}\boldsymbol{A}\delta_{a}-\frac{1}{4\pi c}\boldsymbol{A}\frac{D}{D\boldsymbol{x}}\cdot\left(\frac{1}{c}\partial_{t}\boldsymbol{A}\right)+\frac{1}{4\pi c}\left(\boldsymbol{\nabla}\boldsymbol{A}\right)^{T}\cdot\boldsymbol{\nabla}\varphi\right]+\frac{1}{4\pi c}\boldsymbol{\Omega}:\boldsymbol{A}\boldsymbol{\nabla}\varphi\right\} .\nonumber \\
\label{eq:A_1}
\end{eqnarray}
The second term of Eq.$\thinspace$(\ref{eq:69}) can be written as
\begin{align}
 & \frac{D}{D\boldsymbol{x}}\cdot\left\{ -\frac{D}{Dt}\left[\frac{\partial\mathcal{L}_{KD}}{\partial\boldsymbol{A}_{,t}}\boldsymbol{A}\boldsymbol{x}\right]\right\} :\boldsymbol{\Omega}\nonumber \\
 & =-\frac{D}{D\boldsymbol{x}}\cdot\left\{ \left[\frac{\partial\mathcal{L}_{KD}}{\partial\boldsymbol{A}}\boldsymbol{A}+\frac{\partial\mathcal{L}_{KD}}{\partial\boldsymbol{A}_{,t}}\left(\partial_{t}\boldsymbol{A}\right)-\boldsymbol{\nabla}\cdot\frac{\partial\mathcal{L}_{KD}}{\partial\left(\boldsymbol{\nabla}\boldsymbol{A}\right)}\boldsymbol{A}\right]\boldsymbol{x}\right\} :\boldsymbol{\Omega}\nonumber \\
 & =-\frac{D}{D\boldsymbol{x}}\cdot\left\{ \left[\frac{\partial\mathcal{L}_{KD}}{\partial\boldsymbol{A}}\boldsymbol{A}+\frac{\partial\mathcal{L}_{KD}}{\partial\boldsymbol{A}_{,t}}\left(\partial_{t}\boldsymbol{A}\right)\right]\boldsymbol{x}-\frac{\partial\mathcal{L}_{KD}}{\partial\left(\boldsymbol{\nabla}\boldsymbol{A}\right)}\cdot\boldsymbol{\nabla}\left(\boldsymbol{A}\boldsymbol{x}\right)\right.\nonumber \\
 & \left.\vphantom{\frac{\partial\mathcal{L}_{KD}}{\partial\boldsymbol{A}_{,t}}}-\boldsymbol{\nabla}\cdot\frac{\partial\mathcal{L}_{KD}}{\partial\left(\boldsymbol{\nabla}\boldsymbol{A}\right)}\boldsymbol{A}\boldsymbol{x}+\frac{\partial\mathcal{L}_{KD}}{\partial\left(\boldsymbol{\nabla}\boldsymbol{A}\right)}\cdot\boldsymbol{\nabla}\left(\boldsymbol{A}\boldsymbol{x}\right)\right\} :\boldsymbol{\Omega}\nonumber \\
 & =-\frac{D}{D\boldsymbol{x}}\cdot\left\{ \left[\frac{\partial\mathcal{L}_{KD}}{\partial\boldsymbol{A}}\boldsymbol{A}+\frac{\partial\mathcal{L}_{KD}}{\partial\boldsymbol{A}_{,t}}\left(\partial_{t}\boldsymbol{A}\right)-\frac{\partial\mathcal{L}_{KD}}{\partial\left(\boldsymbol{\nabla}\boldsymbol{A}\right)}\cdot\left(\boldsymbol{\nabla}\boldsymbol{A}\right)\right]\boldsymbol{x}+\frac{\partial\mathcal{L}_{KD}}{\partial\left(\boldsymbol{\nabla}\boldsymbol{A}\right)}\boldsymbol{A}\right.\nonumber \\
 & \left.\vphantom{\frac{\partial\mathcal{L}_{KD}}{\partial\boldsymbol{A}_{,t}}}-\boldsymbol{\nabla}\cdot\frac{\partial\mathcal{L}_{KD}}{\partial\left(\boldsymbol{\nabla}\boldsymbol{A}\right)}\left(\boldsymbol{A}\boldsymbol{x}\right)+\frac{\partial\mathcal{L}_{KD}}{\partial\left(\boldsymbol{\nabla}\boldsymbol{A}\right)}\cdot\boldsymbol{\nabla}\left(\boldsymbol{A}\boldsymbol{x}\right)\right\} :\boldsymbol{\Omega}.\label{eq:A_2}
\end{align}
The EL equations for $\varphi$ and $\boldsymbol{A}$ have been used
in the above derivation. Substituting the Lagrangian density $\mathcal{L}_{KD}$
and $\mathcal{L}_{a}$ in Eq.$\thinspace$(\ref{eq:KD_Lagrangian})
into Eq.$\thinspace$(\ref{eq:A_2}), we have
\begin{alignat}{1}
 & \frac{D}{D\boldsymbol{x}}\cdot\left\{ -\frac{D}{Dt}\left[\frac{\partial\mathcal{L}_{KM}}{\partial\boldsymbol{A}_{,t}}\boldsymbol{A}\boldsymbol{x}\right]\right\} :\boldsymbol{\Omega}\nonumber \\
 & =\frac{D}{D\boldsymbol{x}}\cdot\left\{ \left[-\sum_{a}\frac{q_{a}}{c}\dot{\boldsymbol{X}}_{a}\boldsymbol{A}\delta_{a}-\frac{1}{4\pi c}\boldsymbol{\nabla}\varphi\left(\partial_{t}\boldsymbol{A}\right)+\frac{1}{4\pi}\boldsymbol{\epsilon}:\left[\left(\boldsymbol{\epsilon}:\boldsymbol{\nabla}\boldsymbol{A}\right)\left(\boldsymbol{\nabla}\boldsymbol{A}\right)\right]\right]\boldsymbol{x}:\boldsymbol{\Omega}\right.\nonumber \\
 & \left.\vphantom{\sum_{s,p}}+\frac{1}{4\pi}\boldsymbol{\epsilon}\cdot\left(\boldsymbol{\epsilon}:\boldsymbol{\nabla}\boldsymbol{A}\right)\boldsymbol{A}:\boldsymbol{\Omega}\right\} .\label{eq:A_3}
\end{alignat}
The last two terms in Eq.$\thinspace$(\ref{eq:A_3}) vanish, i.e.,
\begin{align}
 & -\frac{D}{D\boldsymbol{x}}\cdot\left\{ -\boldsymbol{\nabla}\cdot\frac{\partial\mathcal{L}_{KM}}{\partial\left(\boldsymbol{\nabla}\boldsymbol{A}\right)}\left(\boldsymbol{A}\boldsymbol{x}\right)+\frac{\partial\mathcal{L}_{KM}}{\partial\left(\boldsymbol{\nabla}\boldsymbol{A}\right)}\cdot\boldsymbol{\nabla}\left(\boldsymbol{A}\boldsymbol{x}\right)\right\} \nonumber \\
 & =\frac{1}{4\pi}\boldsymbol{\text{\ensuremath{\nabla}}}\cdot\left\{ -\boldsymbol{\nabla}\cdot\left[\boldsymbol{\epsilon}\cdot\left(\boldsymbol{\epsilon}:\boldsymbol{\nabla}\boldsymbol{A}\right)\right]\left(\boldsymbol{A}\boldsymbol{x}\right)+\left[\boldsymbol{\epsilon}\cdot\left(\boldsymbol{\epsilon}:\boldsymbol{\nabla}\boldsymbol{A}\right)\right]\cdot\boldsymbol{\nabla}\left(\boldsymbol{A}\boldsymbol{x}\right)\right\} \nonumber \\
 & =\frac{1}{4\pi}\boldsymbol{\text{\ensuremath{\nabla}}}\cdot\left\{ \boldsymbol{\epsilon}:\left[\boldsymbol{\nabla}\left(\boldsymbol{\epsilon}:\boldsymbol{\nabla}\boldsymbol{A}\right)\right]\left(\boldsymbol{A}\boldsymbol{x}\right)-\boldsymbol{\epsilon}:\left[\left(\boldsymbol{\epsilon}:\boldsymbol{\nabla}\boldsymbol{A}\right)\cdot\boldsymbol{\nabla}\left(\boldsymbol{A}\boldsymbol{x}\right)\right]\right\} \nonumber \\
 & =\frac{1}{4\pi}\boldsymbol{\text{\ensuremath{\nabla}}}\cdot\left\{ \boldsymbol{\nabla}\times\left[\left(\boldsymbol{\epsilon}:\boldsymbol{\nabla}\boldsymbol{A}\right)\right]\boldsymbol{A}\boldsymbol{x}\right\} =0.\label{eq:A_4}
\end{align}
Substituting Eqs.$\thinspace$(\ref{eq:A_1}) and (\ref{eq:A_3})
into Eq.$\thinspace$(\ref{eq:69}), and adding it to Eq.$\thinspace$(\ref{eq:68}),
we obtain Eq.$\thinspace$(\ref{eq:70}).

\bibliographystyle{apsrev}

\bibliography{General_weak_EL_equation}

\end{document}